\documentclass[a4paper,11pt]{article}
\pdfoutput=1
\usepackage[utf8]{inputenc}
\usepackage[T1]{fontenc}
\usepackage{jheppub}
\usepackage{natbib}

\usepackage[usenames,dvipsnames,svgnames,table,x11names]{xcolor}
\usepackage{lmodern,amsmath,multirow,cancel}
\usepackage{booktabs}
\usepackage{xstring}
\usepackage{ifthen}
\usepackage{tabularx} 

\usepackage{ascmac,braket,bm,mathrsfs,amsthm,amsfonts}
\usepackage{hyperref}
\usepackage{graphicx}
\usepackage{subcaption} 
\usepackage{comment}
\usepackage[normalem]{ulem}

\usepackage{titlesec}
\newcommand*{\justifyheading}{\raggedright}
\titleformat{\chapter}[display]
  {\normalfont\huge\bfseries\justifyheading}{\chaptertitlename\ \thechapter}
  {20pt}{\Huge}
\titleformat{\section}
  {\normalfont\Large\bfseries\justifyheading}{\thesection}{1em}{}
\titleformat{\subsection}
  {\normalfont\large\bfseries\justifyheading}{\thesubsection}{1em}{}
\titleformat{\subsubsection}
  {\normalfont\bfseries\justifyheading}{\thesubsubsection}{1em}{}
\usepackage[english]{babel}
\usepackage[dvipsnames]{xcolor}
\newcommand{\tr}{\mbox{tr}}

\numberwithin{equation}{section}

\usepackage{slashed}
\newcommand{\Slash}[1]{{\ooalign{\hfil/\hfil\crcr$#1$}}}

\renewcommand\Im{\mathop{\mathrm{Im}}}

\usepackage{environ}

\newcommand{\ocom}[1]{\textcolor[rgb]{0.32, 0.5, 0.88}{[Osamura: #1]}}

\newcommand{\opn}[1]{\operatorname{#1}}
\NewEnviron{eqsp}{%
\begin{equation}\begin{split}
  \BODY
\end{split}\end{equation}
}
\renewcommand{\proofname}{$\because$}


\usepackage{adjustbox}
\usepackage{tabularx}
\newcolumntype{Y}{>{\centering\arraybackslash}X} 
\usepackage{booktabs} 
\usepackage{colortbl} 
\def\beq#1\eeq{\begin{align}#1\end{align}}

\RequirePackage{xspace}
\def\Bbar    {\kern 0.18em\overline{\kern -0.18em B}{}\xspace}

\preprint{CHIBA-EP-266, IPMU24-0034}

\title{Impact of the Electroweak Weinberg Operator \\on the Electric Dipole Moment of Electron} 

\author[a]{Tatsuya Banno,}
\author[a,b,c]{Junji Hisano,}
\author[b,d]{Teppei Kitahara,}
\author[a]{Kiyoto Ogawa,}
\author[a]{and\\ Naohiro Osamura}

\affiliation[a]{
  Department of Physics, Nagoya University, Furo-cho Chikusa-ku, Nagoya 464-8602 Japan
}
\affiliation[b]{
  Kobayashi-Maskawa Institute for the Origin of Particles and the
  Universe, Nagoya University,
  Furo-cho Chikusa-ku, Nagoya 464-8602 Japan
}
\affiliation[c]{
  Kavli IPMU (WPI), UTIAS, The University of Tokyo, Kashiwa 277-8584, Japan
}

\affiliation[d]{
Department of Physics, Graduate School of Science,
Chiba University, Chiba 263-8522, Japan}

\emailAdd{banno.tatsuya.p8@s.mail.nagoya-u.ac.jp}
\emailAdd{hisano@eken.phys.nagoya-u.ac.jp}
\emailAdd{kitahara@chiba-u.jp}
\emailAdd{ogawa.kiyoto.f8@s.mail.nagoya-u.ac.jp}
\emailAdd{osamura.naohiro.j2@s.mail.nagoya-u.ac.jp}

\abstract{Recent progress in the electric dipole moment (EDM) measurements of the electron using the paramagnetic atom or molecule is remarkable.
In this paper, we calculate a contribution to the electron EDM at three-loop level, introducing the CP-violating Yukawa couplings of new SU(2)$_L$ multiplets. At two-loop level, the Yukawa interactions generate a CP-violating dimension-six operator, composed of three SU(2)$_L$ field strengths, called the electroweak-Weinberg operator.
Another one-loop diagram with this operator inserted induces the electron EDM. We derive the matching condition and find that even if new SU(2)$_L$ particles have masses around the TeV scale, the electron EDM may be larger than the Standard Model (SM) contribution to the paramagnetic atom or molecule EDMs. We also discuss the relation between the Barr-Zee diagram contribution at two-loop level and three-loop one, assuming that the SM Higgs has new Yukawa interactions with the SU(2)$_L$ multiplets. 
}


\begin{document}
\sloppy 

\makeatletter\renewcommand{\@fpheader}{\ }\makeatother

\maketitle

\renewcommand{\thefootnote}{\#\arabic{footnote}}
\setcounter{footnote}{0}


\section{Introduction}

The experimental sensitivities on the electric dipole moment (EDM) of the electron ($d_e$) have been remarkably improved by new technology since the 2010s. 
The Imperial College experiment using YbF molecules surpassed the limits set by atoms, such as Tl, and reached $|d_e|<1.05\times 10^{-27}e\,{\rm cm}$ in 2011 \cite{Hudson:2011zz}. The
bound was quickly updated by the ACME experiment using ThO molecules as $|d_e|<8.7\times 10^{-29}{e}\,{\rm cm}$ in 2013 \cite{ACME:2013pal}, and also reached $1.1\times 10^{-29}{e}\,{\rm cm}$ in 2018 \cite{ACME:2018yjb}. In 2023 \cite{Roussy:2022cmp}, the JILA HfF$^+$ experiment gives the current world record as $|d_e|<4.1\times 10^{-30}e\,{\rm cm}$. It is expected, as reported in Ref.~\cite{Alarcon:2022ero},
that electron EDM will be further improved by several orders of magnitude in the next decades.

The upper bound on the electron EDM gives a severe constraint on physics beyond the Standard Model (BSM), if they have $O(1)$ CP-violating interactions;
the BSM models responsible for the matter-antimatter asymmetry in the Universe are required to have $O(1)$ CP-violating interactions. 
Neutrino oscillation experiments have suggested that
the CP phase in the Pontecorvo-Maki-Nakagawa-Sakata (PMNS) matrix may be $O(1)$ \cite{T2K:2019bcf,T2K:2023smv,NOvA:2023iam}. This might mean that the CP-violating interactions are ubiquitous in BSM models.  The electron EDM is induced by loop diagrams in renormalizable theories with CP-violating interactions,
 and it is proportional to the electron mass ($m_e$) in typical BSM models. Thus, the electron EDM is approximately given as 
\begin{align}
    \frac{d_e}e&\approx \left(\frac{\lambda^2}{16\pi^2}\right)^n\frac{m_e}{\Lambda^2} \sin\phi_{\rm CP}\,,
\end{align}
if the $n$-loop diagrams generate it. If we take the coupling constant $\lambda$ to be comparable to the SU(2)$_L$ and the CP-violating phase $\phi_{\rm CP}$ is $O(1)$, the current electron EDM bound gives the BSM scale $\Lambda$ as $\Lambda \gtrsim 80$\,TeV at the one-loop and 
$\Lambda \gtrsim 4$\,TeV at  two-loop level. In the next decades, the experiments would be sensitive to even the three-loop contribution.

Let us assume SU(2)$_L$ fermions $\psi_{A/B}$ and a scalar $S$ with TeV scale masses have CP-violating Yukawa couplings and discuss the electron EDM induced by the higher-loop diagrams, 
    \begin{equation}
        \mathcal{L} 
        \supset
            -
            \bar{\psi}_B g_{\bar{B} A S} \psi_A S
            -
            \bar{\psi}_A g_{\bar{A} B \bar{S}} \psi_B S^*\,.
        \label{eq:Yukawa intro}
    \end{equation}
 If they are not coupled with the electron at tree level, there is no one-loop contribution to the electron EDM. However, it is known that if the scalar $S$ is identified with the SM Higgs $H$, the electron EDM can be induced by the two-loop Barr-Zee diagrams \cite{Barr:1990vd}. Integrating out the heavy fermions at one-loop level makes the CP-violating $h$-$\gamma$-$\gamma$ coupling, and it induces the electron EDM by another one loop.

    On the other hand, if $S \neq H$, the Yukawa interactions generate a dimension-six CP-violating operator at two-loop level, which contributes to the electron EDM.
    The operator is composed of the SU(2)$_L$ field-strengths, which is given as 
    \begin{align}
        \begin{aligned}
               {\cal L}_{W}
            &=
                - \frac{g^3}{3}
                C_W
                \epsilon^{abc}
                W^a_{\mu\nu}W^{b\nu}_{~\ \rho}\widetilde{W}^{c\rho\mu}
                \,.
            \label{eq:Weinberg operator}
        \end{aligned}
    \end{align}
    Here, $W_{\mu\nu}^a$ ($a=1$--$3$) are the SU(2)$_L$ field-strengths, and  $g$ and $\epsilon^{abc}$ are the SU(2)$_L$ gauge coupling constant and structure constants, respectively. 
    $C_W$ is the Wilson coefficient. 
    The operator in QCD is called the Weinberg operator. 
    Then, we call the operator in Eq.~\eqref{eq:Weinberg operator}  the electroweak-Weinberg operator in this paper. The electroweak-Weinberg operator contributes to the electron EDM by another one loop as $d_e/e\sim (\alpha_2)^2 m_e C_W$  with $\alpha_2=g^2/4\pi$. Overall, it is a three-loop contribution to the electron EDM.

In this paper, we evaluate the electroweak-Weinberg operator contribution to the electron EDM in models with CP-violating Yukawa couplings, given in Eq.~\eqref{eq:Yukawa intro}. 
The Wilson coefficient of the Weinberg operator in QCD is evaluated at two-loop level in general models in Ref.~\cite{Abe:2017sam}. The result is applicable to assess the electroweak-Weinberg operator. However, although the matching condition from the electroweak-Weinberg operator to the electron EDM at the one-loop level has already been evaluated in several papers \cite{Hoogeveen:1987jn,Atwood:1990cm,Boudjema:1990dv,DeRujula:1990db,Novales-Sanchez:2007rsw},  the results are not consistent with each other. Integrating out the extended heavy particles generates the electroweak-Weinberg operator at two-loop level and the CP-violating SU(2)$_L$ dipole operator of the leptons 
at three-loop level in the SMEFT.
The dipole operator also contributes to the electron EDM.
We need to introduce the dimensional regularization for the IR divergence since the $W$ boson is massless in the SMEFT. As a result, the evanescent part of the electroweak-Weinberg operator, which vanishes in the 4-dimensional limit, may also contribute to the electron EDM. 
    We derive the matching conditions of the electroweak and the evanescent operators to the electron EDM. However, in this paper, we use the following matching condition, which includes a contribution from only the electroweak-Weinberg operator, in order to get the numerical results,
    \begin{equation}
        \frac{d_e}{e}
        = 
            \frac{1}{6} \left(\alpha_2\right)^2 m_e C_W \,,
    \end{equation}
    since we do not know the Wilson coefficients of the evanescent operators. We also do not evaluate the Wilson coefficient of the CP-violating SU(2)$_L$ dipole operator of the leptons 
    in the SMEFT since we need to calculate the three-loop diagrams directly, which is out of our scope. Thus, those two contributions are still the uncertainties of $O(1)$ in the prediction of the electron EDM.

 The electron EDM sensitives will be improved in the next decades, as mentioned above. It is expected that even the three-loop contributions will be tested there, as explicitly shown by the numerical analysis in this paper. One of the examples of the tastable models is the SU(2)$_L$ multiplet dark matter models.  
 The electroweak-Weinberg operator is generated if
 the CP-violating Yukawa coupling is introduced to get the dark matter masses in those models. 
If a neutral component in an SU(2)$_L$ multiplet is assumed to be the dark matter in the Universe, its mass is expected to be around TeV scale from the thermal decoupling hypothesis \cite{Cirelli:2005uq,Hisano:2006nn,Cirelli:2007xd,Cirelli:2009uv}. It is challenging to discover such heavy and non-colored particles in LHC experiments, even if the charged components are long-lived.
For example, the current lower bound on the SU(2)$_L$ triplet fermion dark matter mass is at most 660~GeV \cite{ATLAS:2022rme}. 
On the other hand, the SU(2)$_L$ multiplet dark matter may be tested in the next generation experiments of the dark matter direct detection even if they have mass around TeV scale \cite{Hisano:2015rsa, Bloch:2024suj}. 
In addition, the electron EDM induced by the electroweak-Weinberg operator might reach the sensitivities of future electron EDM experiments. From the viewpoint of stability of the dark matter, it is suggested that the fermion SU(2)$_L$ representation is favored to the five-dimensional multiplet ($r=5$) or more, or the scalar is the seven-dimensional one ($r=7$) \cite{Cirelli:2005uq,Cirelli:2007xd,Cirelli:2009uv}; such large SU(2)$_L$ representations automatically lead to an accidental symmetry for the dark matter stability within renormalizable theories. Interestingly, we find that they also lead to an enhancement of the electron EDM, proportional to $r^3$.

This paper is organized as follows. In Sec.~\ref{sec:sun_weinberg}, we review the Weinberg operator in SU($N$) gauge theories induced by the CP-violating Yukawa couplings of the heavier particles.
We follow the results in  Ref.~\cite{Abe:2017sam}, while the analytic formulae are given there.  In Sec.~\ref{sec:eEDM}, we show the contributions to the electron EDM from the electroweak-Weinberg and the evanescent operators. 
In Sec.~\ref{sec:numerical}, the numerical results for 
the contribution of the electroweak-Weinberg operator to the electron EDM are shown. 
We also discuss the relations between the Barr-Zee diagrams and the electroweak-Weinberg operator contributions, assuming the SM Higgs boson has the CP-violating Yukawa coupling. The electroweak-Weinberg operator contributions might be comparable to the radiative correction to the  Barr-Zee diagram contribution. 
Section \ref{sec:conclusion} is devoted to conclusions and discussion.

\section{Weinberg operator in SU($N$) gauge theory at two-loop level}
\label{sec:sun_weinberg}

The Weinberg operator was introduced initially in the SU(3)$_C$.
It can be extended in the SU($N$) gauge theories as 
\begin{equation}
  \mathcal{L}_W 
    = -\frac{g^3}{3}C_Wf^{abc}W^a_{\mu\nu}W^{b\nu}_{~\ \rho}\widetilde{W}^{c\rho\mu}\,.
    \label{eq:WWW coupling}
\end{equation}
Here, $g$, $f^{abc}$, and $W^{a}_{\mu \nu}$ are the gauge coupling constant, the structure constant, and the field strength of SU($N$) gauge theory, respectively,
and its dual is  $\widetilde{W}^{a\mu\nu}=\frac{1}{2}\varepsilon^{\mu\nu\rho\sigma}W^a_{\rho\sigma}$ with $\varepsilon^{0123}=+1$.

\begin{figure}[t]
  \centering
  \includegraphics[width=0.4\linewidth]{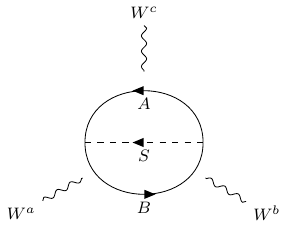}
  \caption{Two-loop diagrams contributing to the SU($N$) Weinberg operator.}
  \label{fig:diagrams_for_Cw}
\end{figure}

The Weinberg operator is generated by the CP-violating Yukawa couplings at two-loop level; see Fig.~\ref{fig:diagrams_for_Cw}.
First, we consider renormalizable Yukawa interactions between fermions ($\psi_A$ and $\psi_B$) and a complex scalar field ($S$) as follows:
\begin{equation}
    \mathcal{L} 
    \supset
        -
        \bar{\psi}_B g_{\bar{B} A S} \psi_A S
        -
        \bar{\psi}_A g_{\bar{A} B \bar{S}} \psi_B S^*\,,
    \label{eq:Yukawa}
\end{equation}
where
\begin{align}
  g_{\bar{B}AS} &= X_{\bar{B}AS}(s+\gamma_5a)\,, \\
  g_{\bar{A}B\bar{S}} &= X_{\bar{A}B\bar{S}}(s^*-\gamma_5a^*)\,.
\end{align}
Here, $X$s are the SU($N$) invariant tensors, 
which depend on the representations of the fields, and $s$ and $a$ are arbitrary complex numbers. 
In Table~\ref{tab:group_factors_SUN}, the representative forms of $X_{\bar{B}AS}$ are listed.

\begin{table}[t]
  \centering
  \renewcommand{\arraystretch}{1.4}
  \scalebox{0.85}{
  \begin{tabular}{cccccc}
  \hline
  $(A,B, S)$ & $\psi_A \psi_B S$ & $X_{\bar{B}AS}$ & $X T_A T_AX^\dagger$ & $XT_A X^\dagger T_B$ & $XX^\dagger T_B T_B$ \\
  \hline
  \hline
  $(N, N, 1)$ & $(\psi_A)_a (\psi_B)_b S$ & $\delta^{ab}$ & $1/2$ & $1/2$ & $1/2$ \\
  $(N, 1, \bar{N})$ & $(\psi_A)_a \psi_B S_i$ & $\delta^{ai}$ & $1/2$ & $0$ & $0$ \\
  \hline
  \hline
  $(\bar{N}, N, {\square\square})$ & $(\psi_A)_a (\psi_B)_b S_{ij}$ & $(\delta^b_i \delta^a_j + \delta^b_j \delta^a_i)/2$ & $(N+1)/4 $  & $ - 1/4 $ & $ (N+1)/4$ \\
  $(N, N, \text{Ad})$ & $(\psi_A)_i (\psi_B)_j S^a $ & $(T_N^a)^j_{~i}$ & $(N^2 -1)/4N$ & $- 1/4N$ & $(N^2 -1)/4N$ \\
  $(N, \text{Ad}, \bar{N})$ & $(\psi_A)_i (\psi_B)^a  S_j$ & $(T_N^a)^j_{~i}$ & $(N^2 -1)/4N$& $N/4$ & $N/2$ \\
  \hline
  \hline
  $(\text{Ad}, \text{Ad}, 1)$ & $(\psi_A)_a (\psi_B)_b  S$ & $\delta^{ab}$ & $N $& $N$ & $N$ \\
  $(\text{Ad}, 1, \text{Ad})$ & $(\psi_A)_a \psi_B  S_i$ & $\delta^{ai}$ & $N $& $0$ & $0$ \\
  $(\text{Ad}, \text{Ad},  \text{Ad})$ & $(\psi_A)_a (\psi_B)_b  S_i$ & $f^{bai}$ & $N^2$& $ f^{a}_{~AA'}f^i_{~A'B'}f^{b}_{~B'B}f^{i}_{~BA}$ & $N^2$ \\
  \hline
  \hline
  \end{tabular}
  }
  \caption{Group factors in typical SU($N$) representations.
  The Young Tableau ${\square\square}$ is the minimal symmetric ($N(N+1)/2$) representation, while $N$ and Ad stand for the fundamental and adjoint representations.}
  \label{tab:group_factors_SUN}
\end{table}

The  Wilson coefficient $C_W$ was evaluated in the SU(3)$_C$ case presented in Ref.~\cite{Abe:2017sam} using the Fock-Schwinger gauge method \cite{Fock:1937dy,Schwinger:1951nm,key86364857,Cronstrom:1980hj,Shifman:1980ui,Dubovikov:1981bf,Novikov:1983gd}.
From the result, we obtain $C_W$ as
\begin{align}
  C_W &= \frac{6}{(4\pi)^4}\Im(sa^*)m_Am_B \nonumber\\
  &\ \times\bigg\{ 
    \left( XT_AT_AX^\dagger \right)g_1(m_A^2,m_B^2,m_S^2) +  \left( XX^\dagger T_BT_B \right)g_1(m_B^2,m_A^2,m_S^2) \nonumber\\
    &\qquad + \left( XT_AX^\dagger T_B \right)\left[ g_2(m_A^2,m_B^2,m_S^2) + g_2(m_B^2,m_A^2,m_S^2) \right]  \bigg\} \label{eq:Wilson coeffiecient}\,.
\end{align}
Here, the group factors are defined as
\begin{align}
  \left( XT_AT_AX^\dagger \right)\delta^{ab} 
  &= \left( X_{\bar{B}AS}(T^a)_{AA'}(T^b)_{A'A''}X^\dagger_{\bar{A}''B\bar{S}} \right)\,, \\
  \left( XX^\dagger T_BT_B \right)\delta^{ab} 
  &= \left( X_{\bar{B}AS}X^\dagger_{\bar{A}B''\bar{S}}(T^a)_{B'B''}(T^b)_{B''B} \right)\,, \\
  \left( XT_AX^\dagger T_B \right)\delta^{ab} 
  &= \left( X_{\bar{B}AS}(T^a)_{AA'}X^\dagger_{\bar{A}'B'\bar{S}}(T^b)_{B'B}  \right)\,,
\end{align}
and the subscript of $T^a$, such as $A$ or $B$, denotes that it is the generator for the fermion field $A$ or $B$ with $a = 1,2, \cdots, N^2-1$.
In Table~\ref{tab:group_factors_SUN},
we show the explicit values of the group factors for 
typical SU$(N)$ representations. 
The two-loop functions $g_1$ and $g_2$ are defined as\footnote{
Here, $g_1=-f_1$ and $g_2=-f_2$ where $f_1$ and $f_2$ are defined in Ref.~\cite{Abe:2017sam}.
}
\begin{align}
  g_1(x_1,x_2,x_3) &= \left( 2\bar{I}_{(4;1)} + 4x_1\bar{I}_{(5;1)} \right)(x_1;x_2;x_3)\,, \\
  g_2(x_1,x_2,x_3) &= \left( \bar{I}_{(3;2)} + x_1\bar{I}_{(4;2)} \right)(x_1;x_2;x_3)\,,
\end{align}
where $\bar{I}_{(n;m)}$ is the UV finite two-loop function, which is given as
\begin{align}
  \bar{I}_{(n;m)}(x_1;x_2;x_3)
  = \frac{1}{(n-1)!(m-1)!} \frac{d^{n-1}}{dx_1^{n-1}} \frac{d^{m-1}}{dx_2^{m-1}} \bar{I}(x_1;x_2;x_3)\,,
\end{align}
and the explicit form of $\bar{I}$ is
\begin{align}
  & \bar{I}(x_1;x_2;x_3)
  =
  -\frac12\biggl[(-x_1+x_2+x_3)\log\frac{x_2}{Q^2}\log\frac{x_3}{Q^2}
+(x_1-x_2+x_3)\log\frac{x_1}{Q^2}\log\frac{x_3}{Q^2}
\nonumber\\
&\quad 
+(x_1+x_2-x_3)\log\frac{x_1}{Q^2}\log\frac{x_2}{Q^2}
-4\left(x_1\log\frac{x_1}{Q^2}
+x_2\log\frac{x_2}{Q^2}
+x_3\log\frac{x_3}{Q^2}\right)
\nonumber\\
&\quad 
+5(x_1+x_2+x_3)+\xi(x_1,x_2,x_3)\biggr]\,,
\end{align}
and
\begin{align}
& \xi(x_1,x_2,x_3)=
R\left[2 \log\left(\frac{x_3+x_1-x_2-R}{2 x_3}\right)\log\left(\frac{x_3-x_1+x_2-R}{2 x_3}\right)
-\log\frac{x_1}{x_3}\log\frac{x_2}{x_3}
\right.\nonumber\\
&
\quad  \left.
-2{\rm Li}_2\left(\frac{x_3+x_1-x_2-R}{2 x_3}\right)
-2{\rm Li}_2\left(\frac{x_3-x_1+x_2-R}{2 x_3}\right)
+\frac{\pi^2}{3}
\right]\,,
\end{align}
where $R=\sqrt{x_1^2+x_2^2+x_3^2-2x_1x_2-2x_2x_3-2x_3x_1}$ and $Q^2=4\pi\mu^2e^{-\gamma_E}$\cite{Ford:1992pn,Espinosa:2000df,Martin:2001vx}. 
Here, since the subdiagrams contain UV-divergence, we introduce the dimensional regularization only in the loop momentum integral ($d=4-2\epsilon$ and  $\mu$ is the renormalization scale).
However, the final result is UV-finite and independent of $\mu$. In Ref.~\cite{Abe:2017sam}, $g_1$ and $g_2$ are numerically evaluated. Since we use the analytic mass function for the two-loop vacuum diagram, $\bar{I}(x_1;x_2;x_3)$, we can evaluate them analytically. For more details of the two-loop functions, see Ref.~\cite{Hisano:2023izx}.

The analytic formula for $C_W$ is useful in the discussion of the behavior of heavy particle decoupling. When $m_A \ll m_B$ and $m_S$, the following effective theory approach works in the evaluation of the SU($N$) Weinberg operator.
First, the SU($N$) EDM of the lighter fermion $\psi_A$ is generated at one-loop level by 
integrating $\psi_B$ and $S$ out, as follows:
\begin{align}
\begin{aligned}
  d_A^N
  = &-\frac{1}{(4\pi)^2}\Im(sa^*)(X^\dagger T_SX)\frac{m_B}{m_S^2}f_S(x_{BS}) \\
  &-\frac{1}{(4\pi)^2}\Im(sa^*)(X^\dagger T_BX)\frac{m_B}{m_S^2}f_B(x_{BS})\,,
  \end{aligned}
\end{align}
where $x_{BS}= m_B^2/m_S^2$ and
\begin{align}
  f_S(x) &= \frac{1-x^2+2x\log x}{(1-x)^3}\,, \\
  f_B(x) &= -\frac{3-4x+x^2+2\log x}{(1-x)^3}\,.
  \label{eq:fb}
\end{align}
The group factors are given as 
\begin{align}
(X^\dagger T_SX) (T^a)_{AA'} &=
  \left( X^\dagger_{\bar{A}B S}(T^a)_{SS'}X_{\bar{B}A'S'} \right)\,, \\
(X^\dagger T_BX) (T^a)_{AA'} &=
   \left( X^\dagger_{\bar{A}B S}(T^a)_{BB'}X_{\bar{B}'A'S}\right)\,.
\end{align}

Next when $\psi_A$ is decoupled, 
the SU($N$) Weinberg operator is induced at one-loop level 
from the operator of $d_A^N$, as
\begin{align}
  C_W = \frac{1}{(4\pi)^2}N(r_A)\frac{d_A^N}{m_A}\,. \label{eq: d_A_to_C_W}
\end{align}
where the Dynkin index $N(r)$ is defined by ${\rm tr}(T^aT^b)=N(r)\delta^{ab}$.
This is directly derived from the analytical formula in Eq.~\eqref{eq:Wilson coeffiecient} by taking $m_A \ll m_B$ and $m_S$. In this case, the mass functions are approximately given as
\begin{align}
  g_1(m_A^2,m_B^2,m_S^2)
&\simeq  \frac{1}{6m_A^2m_S^2}f_S(x_{BS})\,, \\
  g_2(m_A^2,m_B^2,m_S^2)
&\simeq -\frac{1}{6m_A^2m_S^2}\left[f_S(x_{BS})+f_B(x_{BS})\right]\,.
\end{align}
In $C_W$, $g_{1/2}(m_B^2,m_A^2,m_S^2)$ are negligible compared with $g_{1/2}(m_A^2,m_B^2,m_S^2)$.
Since $X$ is an invariant tensor in SU($N$), we get \cite{Abe:2017sam}
\begin{align}
  XT_AT_AX^\dagger&=N(r_A)( X^\dagger T_BX -X^\dagger T_S X) \,, \\
  XT_AX^\dagger T_B&= N(r_A) X^\dagger T_BX\,.
\end{align}
Using the above formulae, we can derive Eq.~\eqref{eq: d_A_to_C_W}. From the above exercise, we found that when one of the fermions is lighter, the Weinberg operator is enhanced by the lighter mass, a factor of $m_B/m_A$ in this case.

When $m_S\gg m_B>m_A$, integration 
of $S$ generates SU($N$) EDMs for $\psi_A$ and $\psi_B$ in addition to the CP-violating four-Fermi operator of $\psi_A$ and $\psi_B$ \cite{Banno:2023yrd}. The SU($N$) EDMs are proportional to $m_B/m_S^2$ and $m_A/m_S^2$, respectively. The SU($N$) EDMs and the four-Fermi operator are mixed due to their anomalous dimensions \cite{Hisano:2012cc}. Integration of $\psi_A$ and $\psi_B$
generates the Weinberg operator proportional to the physical SU($N$) EDMs, as explained above. Then, the dominant contribution is proportional to $(m_B/m_A)/m_S^2\log(m_S^2/m_B^2)$ when $m_S\gg m_B>m_A$. The logarithmic enhancement appears in Eq.~\eqref{eq:fb}.

In this section, we discuss the Weinberg operator in SU($N$) gauge theories. We assumed that the SU($N$) gauge-boson mass is negligible in the two-loop contributions to the Weinberg operator in Eq.~\eqref{eq:Wilson coeffiecient}. In the next section, we consider the electroweak-Weinberg operator and apply the above results to the evaluation of the electron EDM. This implies that the BSM particles in the CP-violating Yukawa couplings are heavy enough for the $W$ boson to be negligible. 

For later convenience, in Table~\ref{tab:group_factors_SU2}, we listed the explicit group factors in the SU(2)$_L$ representations. 
We also obtain the general formulae of the group factors for the $(A,B,S) = (r,r,1)$ and $(r,1,r)$ representations.
They correspond to
the Dynkin index $N(r) = r(r^2 -1)/12$ which can be obtained from the quadratic Casimir operator $C_2(r) \delta_{ij} (\equiv (T^a T^a)_{ij}) = ((r^2-1)/4) \delta_{ij}$.
Notably, it is important for this paper that the group factors are enhanced by the cubic power of $r$, which amplifies the electroweak-Weinberg operator.

\begin{table}[t]
\centering
\renewcommand{\arraystretch}{1.3}
\begin{tabular}{cccccc}
\hline
$(A,B, S)$ & $\psi_A \psi_B S$ & $X_{\bar{B}AS}$ & $X T_A T_AX^\dagger$ & $XT_A X^\dagger T_B$ & $XX^\dagger T_B T_B$ \\
\hline
\hline
$(2, 2, 1)$ & $(\psi_A)_a (\psi_B)_b S$ & $\delta^{ab}$ & $1/2$ & $1/2$ & $1/2$ \\
$(2, 1, 2)$ & $(\psi_A)_a \psi_B S_i$ & $\delta^{ai}$ & $1/2$ & $0$ & $0$ \\
$(3, 3, 1)$ & $(\psi_A)_{\alpha} (\psi_B)_{\beta} S$ & $\delta^{\alpha\beta}$ & $2$ & $2$ & $2$ \\
$(3, 1, 3)$ & $(\psi_A)_{\alpha} \psi_B S_{\gamma}$ & $\delta^{\alpha \gamma} $ & $2$ & $0$ & $0$ \\
$(3, 2, 2)$ & $(\psi_A)_\alpha (\psi_B)_b S_{i}$ & $(T^\alpha)^{bi}$ & $1$  & $1/2$ & $3/8$ \\
$(2, 2, 3)$ & $(\psi_A)_a (\psi_B)_b S_{\gamma}$ & $(T^\gamma)^{ba}$ & $3/8$  & $-1/8$ & $3/8$ \\
$(3, 3, 3)$ & $(\psi_A)_\alpha (\psi_B)_\beta S_\gamma$ & $\epsilon^{\beta\alpha \gamma}$ & $4$ & $2$ & $4$ \\
\hline
\hline
$(r, r, 1)$ & $(\psi_A)_{a_r} (\psi_B)_{b_r} S$ & $\delta^{a_r b_r}$ & $ r (r^2-1)/12$  & $r (r^2-1)/12 $ & $ r (r^2-1)/12$ \\
$(r, 1, r)$ & $(\psi_A)_{a_r} \psi_B S_{i_r}$ & $\delta^{a_r i_r}$ & $ r (r^2-1)/12$ & $0$ & $0$ \\
\hline
\hline
\end{tabular}
\caption{Group factors in the SU(2)$_L$ representations.
The last two rows show the simple cases of the $r$-dimensional representations.}
\label{tab:group_factors_SU2}
\end{table}

\section{Electron EDM induced by the  electroweak-Weinberg operator}
\label{sec:eEDM}

It was already discussed earlier in Refs.~\cite{Hoogeveen:1987jn,Atwood:1990cm,Boudjema:1990dv,DeRujula:1990db,Novales-Sanchez:2007rsw} that the nonzero Wilson coefficient of the electroweak-Weinberg operator in Eq.~\eqref{eq:Wilson coeffiecient} generates the electron EDM through a one-loop diagram. While the one-loop diagram is finite, it is found that the matching condition depends on the UV regularization schemes, such as the dimensional, the Pauli-Villars, and the momentum cut-off regularizations \cite{Boudjema:1990dv}. The results in 
Refs.~\cite{Hoogeveen:1987jn,Atwood:1990cm,Boudjema:1990dv,DeRujula:1990db,Novales-Sanchez:2007rsw} differ from  the recent evaluation in Ref.~\cite{Dekens:2019ept} while adopting the dimensional regularization.

In order to understand the origin of the regularization, first we review the derivation of the electron EDM from the full UV theories. The SMEFT, which includes the higher-dimensional operators, is derived by integrating out the heavy particles in BSM in a limit of vanishing the Higgs vacuum expectation value. In our models, the electroweak-Weinberg operator and also the CP-violating SU(2)$_L$ dipole operator of the leptons, $\bar{l} T_a W^{a}_{\mu\nu}  \sigma^{\mu\nu}\gamma_5 e H$, are relevant to us ($l$, $e$, and $H$ are for doublet and singlet leptons and the doublet Higgs in the SM, respectively). The former is generated at two-loop level, while the latter is at three-loop level. The latter operator also contributes to the electron EDM after the Higgs gets the vacuum expectation value.  When evaluating the Wilson coefficients for those operators, we have to introduce the regularization even if they are UV finite. The IR divergence may appear in the evaluation of the dipole operator, though the coefficient is finite. Then, when evaluating the electron EDM from the SMEFT, we have to introduce the UV regularization in a consistent way with the IR regularization.

The dimensional regularization is suitable for regularizing the IR and UV divergences. However, since the Lagrangian is promoted to be $d$-dimensional,  we cannot ignore the evanescent operators for the electroweak-Weinberg operator, which vanishes in a limit of $d\rightarrow 4$. After including the evanescent operators, the $W^+$-$W^-$-$\gamma$ interactions in the electroweak-Weinberg operator are given  as  
    \begin{eqsp}
        {\cal L}_{W}
        &=
            - \frac{2 i e g^2}{3}
            C_{W}
            \left[
                \bar{W}^{-}_{\mu \nu}
                \bar{W}^{+\nu}_{\quad\lambda}
                \widetilde{\bar{F}}^{\lambda \mu}
                +
                \bar{F}_{\mu \nu}
                \left(
                    \bar{W}^{-\nu}_{\quad\lambda}
                    \widetilde{\bar{W}}^{+\lambda \mu}
                    -
                    \bar{W}^{+\nu}_{\quad\lambda}
                    \widetilde{\bar{W}}^{-\lambda \mu}
                \right)
            \right]
            \\
        &\quad
            - \frac{2 i e g^2}{3}
            C_{W^{(d-4)}}
            \left[
                W^{-}_{\bar{\mu} \hat{\nu}}
                W^{+\hat{\nu}}_{\quad\bar{\lambda}}
                \widetilde{\bar{F}}^{\lambda \mu}
                +
                F_{\bar{\mu} \hat{\nu}}
                \left(
                    W^{-\hat{\nu}}_{\quad\bar{\lambda}}
                    \widetilde{\bar{W}}^{+\lambda \mu}
                    -
                    W^{+\hat{\nu}}_{\quad\bar{\lambda}}
                    \widetilde{\bar{W}}^{-\lambda \mu}
                \right)
            \right]
            +
            \cdots  
            \\
        &\equiv
            C_W \left(\bar{O}_1 + \bar{O}_2\right)
            +
            C_{W^{(d-4)}} \left(\hat{O}_1 + \hat{O}_2\right)
            +
            \cdots \,,
            \label{eq:ddimEWWeinberg}
    \end{eqsp}
where $W^+$-$W^-$-$Z$ interactions are omitted.
Here, $W_{\mu\nu}^\pm$ and $F_{\mu\nu}$ the field strengths of $W^\pm$ boson and photon, respectively.
    Bars and hats on field strengths (metrics, $\gamma$ matrices,  momentums, and polarization vectors in the below) conventionally represent $4$ and $(d-4)$-dimension, respectively. We employ the BMHV scheme in this section, and then the definition of the Levi-Civita $\varepsilon$ tensor is not changed \cite{tHooft:1972tcz,Breitenlohner:1977hr}.
    We parameterize Wilson coefficients of the evanescent operators as $C_{W^{(d-4)}}$ because  $C_{W^{(d-4)}}$ is not necessarily equivalent to $C_W$ due to violation of the $d$-dimensional Lorentz symmetry in the BMHV scheme.

    \begin{figure}
        \begin{center}
                \includegraphics[width = 0.4\textwidth]{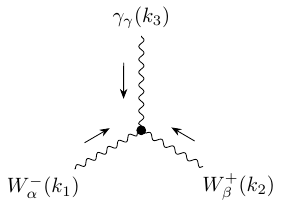}
                \quad 
                \includegraphics[width = 0.45 \textwidth]{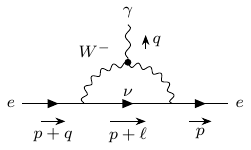}
            \caption{
            (Left) The momentum assignment for the Feynman rule of the electroweak-Weinberg operator. 
            (Right) The one-loop diagram for the electron EDM induced by the electroweak-Weinberg operator. 
             \label{fig:EDM through CW}}
        \end{center}
    \end{figure}

We found that Refs.~\cite{Hoogeveen:1987jn,Atwood:1990cm,Boudjema:1990dv,DeRujula:1990db,Novales-Sanchez:2007rsw} did not argue the contribution from the evanescent operators enough, so the result is unreliable.
In this section, we explicitly calculate each contribution from 
the electroweak-Weinberg operator and the evanescent operators. Notice that 
$\bar{O}_2$ satisfies $\bar{O}_2 = 2 \bar{O}_1$ by the identity of the Levi-Civita $\varepsilon$ tensor. Moreover, the contribution to the electron EDM from
$\hat{O}_2$ is ignored, because the operator is  proportional to $F_{\bar{\mu}\hat{\nu}}$ which contains the $(d-4)$-dimensional Lorentz index or momentum of the external photon. Thus, now we evaluate the contribution from $\bar{O}_1$ and  $\hat{O}_1$.

The Feynman rules for these operators are expressed as
    \begin{align}
        & \begin{aligned}
            &
                i \Gamma_{\bar{O}_1}
                \left[
                    W_\alpha^{-}\left(k_1\right), 
                    W_\beta^{+}\left(k_2\right), 
                    \gamma_\gamma\left(k_3\right)
                \right]
                \\
            &\quad =
                \frac{2}{3} i e g^2 C_{W}
                \bar{\varepsilon}^{\rho \mu \kappa  \delta}
                \left(
                    \bar{k}_{1, \mu} \bar{g}_{\alpha \nu}
                    -
                    \bar{k}_{1, \nu} \bar{g}_{\alpha \mu}
                \right)
                \left(
                    \bar{k}_2^\nu \bar{g}_{\beta \rho}
                    -
                    \bar{k}_{2, \rho} \bar{g}_\beta^\nu
                \right) 
                \bar{k}_{3, \kappa} \bar{g}_{\gamma \delta} \,,
        \end{aligned}
        \\
        & \begin{aligned}
            &
                i \Gamma_{\hat{O}_1}
                \left[
                    W_\alpha^{-}\left(k_1\right), 
                    W_\beta^{+}\left(k_2\right), 
                    \gamma_\gamma\left(k_3\right)
                \right]
                \\
            &\quad =
                \frac{2}{3} i e g^2 C_{W^{(d-4)}}
                \bar{\varepsilon}^{\rho \mu \kappa  \delta}
                \left(
                    \bar{k}_{1, \mu} \hat{g}_{\alpha \nu}
                    -
                    \hat{k}_{1, \nu} \bar{g}_{\alpha \mu}
                \right)
                \left(
                    \hat{k}_2^\nu \bar{g}_{\beta \rho}
                    -
                    \bar{k}_{2, \rho} \hat{g}_\beta^\nu
                \right) 
                \bar{k}_{3, \kappa} \bar{g}_{\gamma \delta} \,.
        \end{aligned}
    \end{align}
    The momentum assignment is shown on the left of Fig.~\ref{fig:EDM through CW}.       
    The one-loop amplitude to the electron EDM from Eq.~\eqref{eq:ddimEWWeinberg} shown in the right of Fig.~\ref{fig:EDM through CW} is computed as
    \begin{eqsp}
        i\mathcal{M}&= \frac{eg^4}{3} \bar{\varepsilon}^{\rho\mu\kappa\delta}
        \int\frac{d^d \ell}{(2\pi)^d}\frac{\bar{u}(p) \gamma^\beta P_L (\Slash{\ell} + \Slash{p}) \gamma^\alpha P_L u(p+q)}{(\ell + p)^2(\ell^2 - m_W^2)}
        \bigg[
            3 C_{W} (-\bar{\ell}_\mu \bar{g}_{\alpha\nu} + \bar{\ell}_\nu \bar{g}_{\alpha\mu})(\bar{\ell}^\nu \bar{g}_{\beta\rho} - \bar{\ell}_\rho \bar{g}^\nu_\beta) \\
            &\quad  + C_{W^{(d-4)}}(-\bar{\ell}_\mu \hat{g}_{\alpha\nu} + \hat{\ell}_\nu \bar{g}_{\alpha\mu})(\hat{\ell}^\nu \bar{g}_{\beta\rho} - \bar{\ell}_\rho \hat{g}^\nu_\beta)
        \bigg]
         \bar{q}_\kappa \bar{g}_{\gamma\delta} + \mathcal{O}(q^2) \\
        &\simeq \frac{eg^4}{3} \bar{q}_\kappa\
        \bigg[ 
            3 C_{W}\left\{(\tr[\bar{g}]+2)A_3 + (\tr[\bar{g}]-2)A_2 \right\} + C_{W^{(d-4)}} \tr[\hat{g}](A_3 + A_2) 
        \bigg]\\
                &\quad \times
        \bar{u}(p)\bar{\gamma}_\rho\bar{\Slash{p}}\bar{\gamma}_\mu\bar{\varepsilon}^{\rho\mu\kappa\delta}P_Lu(p+q) \bar{g}_{\gamma \delta} \ + \cdots \\
        &= -\frac{eg^4}{3}m_e \bar{q}^\kappa
        \bigg[ 
            3 C_{W}\left\{(\tr[\bar{g}]+2)A_3 + (\tr[\bar{g}]-2)A_2 \right\} + C_{W^{(d-4)}} \tr[\hat{g}](A_3 + A_2) 
        \bigg]
        \\
        &\quad \times \bar{u}(p)\bar{\sigma}_{\kappa\gamma}\gamma_5u(p+q) 
         + \cdots\,,    \label{eq:loopexactO1} 
    \end{eqsp}
    where $\tr[\bar{g}]=4$ and $\tr[\hat{g}]=-2\epsilon$ with $d=4-2\epsilon$. Here, we neglect the neutrino mass and the PMNS matrix. 
    The last line is obtained from the following identity derived by the equation of motions in $4$-dimension,
    \begin{equation}
        \bar{u}(p) 
        \bar{\gamma}_\nu \bar{\Slash p} \bar{\gamma}_\mu 
        \bar{\varepsilon}^{\mu \nu \rho \sigma} 
        P_L u(p+q) \bar{q}_\rho \bar{\varepsilon}_\sigma
        =
            m_e 
            \bar{u}(p) 
            \bar{\sigma}^{\rho \sigma} \gamma_5 u(p+q) \bar{q}_\rho 
            \bar{\varepsilon}_\sigma \,,
    \end{equation}   
    with $\bar{\sigma}^{\mu \nu}=\frac{i}{2}\left[\bar{\gamma}^\mu, \bar{\gamma}^\nu\right]$ and the $4$-dimensional photon's polarization vector $\bar{\varepsilon}_\sigma$.
    In Eq.~\eqref{eq:loopexactO1}, we define the following $d$-dimensional loop functions up to $O (p^2 / m_W^2)$, 
    \begin{align}
        & \mu^{4-d}
        \int \frac{d^d \ell}{(2 \pi)^d} \frac{\ell^\mu \ell^\nu}{(\ell+p)^2(\ell^2 - m_W^2)^2}
        =
            A_2 g^{\mu \nu}
            \,,
            \\
        &\mu^{4-d}
        \int \frac{d^d \ell}{(2 \pi)^d} 
        \frac{
            \ell^{\mu_1} \ell^{\mu_2} \ell^{\mu_3}
            }{
                (\ell+p)^2 (\ell^2 - m_W^2)^2
            }
        =
            A_3 
            \left(
                p^{\mu_1} g^{\mu_2 \mu_3}
                +
                p^{\mu_2} g^{\mu_3 \mu_1}
                +
                p^{\mu_3} g^{\mu_1 \mu_2}
            \right)
            \,,
    \end{align}
    with
    \begin{align}
        A_2
        &=
            \frac{i}{(4\pi)^2}
            \frac{1}{4}
            \left(
                \frac{1}{\epsilon}
                +
                \ln \frac{Q^2}{m_W^2}
                +
                \frac{1}{2}
            \right)\,,\\        
        A_3
        &=
            - \frac{i}{(4\pi)^2}\frac{1}{12}
            \left( 
                \frac{1}{\epsilon}
                +
                \log \frac{Q^2}{m_W^2}
                +
                \frac{5}{6}
            \right)\,.
    \end{align}
    
    Comparing the dipole moment amplitude to the definition of the electron EDM operator $\mathcal{L}_{\text {eff}} =
    -\frac{i}{2} d_e \bar{e}(\sigma \cdot F) \gamma_5 e$, 
     we obtain the one-loop matching condition as
    \begin{eqsp}
        \frac{d_e^{C_W}}{e} 
       &= 
            i \frac{g^4}{3} m_e
            \left[
                3 C_{W}
                \left\{
                    (\opn{tr} [\bar{g}] + 2) A_3
                    +
                    (\opn{tr} [\bar{g}] -2) A_2
                \right\}
                +
                C_{W^{(d-4)}}
                \opn{tr} [\hat{g}]
                \left( A_3 + A_2 \right)
            \right]
            {+
            \mathcal{O} \left(\frac{m_e^2}{ m_W^2}\right)}
            \\
        &=
            \left( 
                \frac{1}{6} C_{W}
                +
                \frac{1}{9} C_{W^{(d-4)}}
            \right)
            \left(\alpha_2\right)^2   m_e 
            C_W 
            +
            \mathcal{O} \left(\frac{m_e^2}{ m_W^2}\right)\,.
    \end{eqsp}
    This is consistent with the result with Refs.~\cite{Dekens:2019ept,Abe:2024mwa} 
    though they assume $C_{W^{(d-4)}}$ is equal to $C_{W}$.\footnote{
     Here, we adopt the BMHV scheme for this derivation, but Ref.~\cite{Dekens:2019ept} mentions that this result is also produced by the Naive-dimensional-regularization method.    %
    }
    On the other hand, in Ref.~\cite{Boudjema:1990dv}, ${\cal L}= 3 C_W(\bar{O}_1 + \hat{O}_1)$ was considered, so that they derived the matching condition $d_e / e = \frac{1}{2} \left(\alpha_2\right)^2 m_e C_W$. (They implicitly assumed $\hat{O}_2=2 \hat{O}_1$.)
    We summarize the contributions to the electron EDM from $\bar{O}_1$, $\bar{O}_2$, $\hat{O}_1$, and $\hat{O}_2$ in Table~\ref{tab:de summary}.

    \begin{table}[t]
        \centering
        \renewcommand{\arraystretch}{1.3}
        \begin{tabular}{cccccc}
        \hline
             & &\multicolumn{2}{c}{$4$-dim.~~~} & \multicolumn{2}{c}{ $(d-4)$-dim.} \\
              \hline 
              \hline
            $ - i e g^2 \frac{2}{3}  {W}^{-}{W}^+\widetilde{{F}}$ & & $\bar{O}_1: $ & $(1/18) C_{W}$ & 
             $\hat{O}_1:$&$  (1/9) C_{W^{(d-4)}}$\\ 
             $- i e g^2 \frac{2 }{3} F ( W^-\widetilde{{W}}^+ - W^+\widetilde{{W}}^-)$ & & $\bar{O}_2:$&$  (1/18 + 1/18) C_{W}$ & $\hat{O}_2:$&$  0$ 
            \\ 
            \hline 
            \hline
            total & & \multicolumn{2}{c}{$(1/6) C_{W}$~~}& \multicolumn{2}{c}{$(1/9) C_{W^{(d-4)}}$ }\\
            \hline
            \hline 
        \end{tabular}
        \caption{Summary of contributions from each operator to the electron EDM, which is normalized by $1/[\left(\alpha_2\right)^2 m_e]$.}
        \label{tab:de summary}
    \end{table}

    Now, we derive the matching condition to the electron EDM, though there are two uncertainties in the evaluation of the electron EDM: the contributions from the evanescent operators and the threshold correction to the electron EDM, as we mentioned in the head of this section.
    The latter uncertainty is removed only by the three-loop calculation. For the former one,  
    we may need to evaluate the Wilson coefficients of the evanescent operators, for example, in the BMHV scheme. Those tasks are out of our current scope. 
    In the below numerical analysis, we adopt the following matching condition from only $\bar{O}_1$ and $\bar{O}_2$,
    \begin{equation}
        \frac{d_e^{C_W}}{e}
        =
            \frac{1}{6}
            \left(\alpha_2\right)^2 m_e 
            C_W 
            +
            \mathcal{O} \left(\frac{m_e^2}{m_W^2}\right)\,.
    \end{equation}

\section{Numerical analysis}
\label{sec:numerical}

In this section, we evaluate the electron EDM contribution from the electroweak-Weinberg operator in Eq.~\eqref{eq:Weinberg operator} induced by  the CP-violating Yukawa interactions in  Eq.~\eqref{eq:Yukawa},
and discuss the prospects for future experiments. 
First, we introduce the BSM scalar $S$ and fermions $\psi_A$ and $\psi_B$. Then, the scalar is not the SM Higgs ($S\ne H$).
In this case, if the BSM particles do not have any Yukawa interaction with the matter fields in the SM sector, the electroweak-Weinberg operator may be the leading contribution to the electron EDM, although it is generated at three-loop level. Next, we also consider a scenario in which the scalar $S$ is identified with the SM Higgs ($S = H$). 
Here, we assume that the BSM fermions do not have any Yukawa interaction with the matter fields in the SM sector.
In this scenario, the electron EDM receives another contribution from the Barr-Zee diagrams at two-loop level \cite{Barr:1990vd,Hisano:2014kua,Nagata:2014wma}, and the electron EDM is expected to be dominated by the two-loop contributions. In addition, the next-to-leading order (NLO) contribution to the Barr-Zee diagrams is evaluated in Ref.~\cite{Kuramoto:2019yvj}, which are three-loop level.
     Therefore, we will compare the contributions from the electroweak-Weinberg operator and the NLO to the Barr-Zee diagrams, both of which are three-loop orders.

\subsection{$S \neq H$}
Here, we consider the case of the scalar $S$ not being the SM Higgs.
The electroweak-Weinberg operator induced at two-loop level depends on the SU(2)$_L$ representations of the scalar $S$ and fermions $\psi_A$ and $\psi_B$ in the Yukawa couplings. First, we consider  scenarios of 
$(A,B,S) = (r,r,1)$ and $(r,1,r)$, where $r$ means an $r$-dimensional multiplet in the SU(2)$_L$ gauge group. 
The SU(2)$_L$ group factors in Eq.~\eqref{eq:Wilson coeffiecient} are listed in Table~\ref{tab:group_factors_SU2}.
For simplicity, we assume two particle masses are common, such as (1) $m_B = m_S$, (2) $m_A=m_S$, and (3)  $m_A = m_B$.  In those mass spectra, when the two masses are heavier than another one, the Wilson coefficient $C_W$ is expressed as\footnote{
Here, we expand the mass functions where $x < y = z$ as,
        \begin{align}
            & g_1 (x,y,z)
            \simeq
                \frac{1}{18 x y} \,,
                \nonumber\\
            & g_1 (y,x,z)
            \simeq
                \frac{1}{y^2} \,,
                \nonumber\\
            & g_2 (x,y,z) + g_2 (y,x,z)
            \simeq
                - \frac{1}{6 x y} \,,
        \end{align}
        while in $z < x < y$ case,
        \begin{align}
            g_1(x,y,z)
            &\simeq
                \frac{1}{6xy}
                +
                \frac{3+2\log\frac{x}{y}}{6y^2} \,,
                \nonumber\\
            g_1 (y,x,z)
            &\simeq
                \frac{1}{6 y^2} \,,
                \nonumber\\
            g_2(x,y,z)
            &\simeq
                -\frac{1}{3xy}
                -
                \frac{4+3\log{\frac{x}{y}}}{6y^2} \,,
                \nonumber\\
            g_2(y,x,z)
            &\simeq 
                \frac{1}{6y^2} \log{\frac{x}{y}}\,.
\label{eq:appriximationofg1andg2}
        \end{align}
        We checked these expansions analytically and numerically.}
        \begin{align}
            &\begin{aligned}
                C_W^{(r,r,1)}
                &=
                    \frac{1}{2 (4 \pi)^2}
                    \opn{Im} (s a^*)
                    m_A m_B
                    r (r^2 - 1)
                    \\
                &\qquad \times
                    \left\{
                        g_1 (m_A^2, m_B^2, m_S^2)
                        +
                        g_1 (m_B^2, m_A^2, m_S^2)
                        +
                        g_2 (m_A^2, m_B^2, m_S^2)
                        +
                        g_2 (m_B^2, m_A^2, m_S^2)
                    \right\}
                    \\
                &\simeq
                    \left\{
                        \begin{aligned}
                            &
                                \frac{1}{18 (4 \pi)^2}
                                \opn{Im} (s a^*)
                                r (r^2 - 1)
                                \frac{1}{m_A m_B} 
                                \qquad
                            &
                                (m_A < m_B = m_S ) \,, 
                                \\
                            &
                                -
                                \frac{1}{12 (4 \pi)^2}
                                \opn{Im} (s a^*)
                                r (r^2 - 1)
                                \frac{1}{m_A m_B}
                                \quad 
                            &
                                ( m_S < m_A = m_B ) \,,
                        \end{aligned}
                    \right.
                \label{eq:Weinberg rr1}
            \end{aligned}
            \\
            &\begin{aligned}
                C_W^{(r,1,r)}
                &=
                    \frac{1}{2 (4 \pi)^2}
                    \opn{Im} (s a^*)
                    m_A m_B
                    r (r^2 - 1)
                    g_1 (m_A^2, m_B^2, m_S^2)
                    \\
                &\simeq
                    \left\{
                        \begin{aligned}
                            &
                                \frac{1}{36 (4 \pi)^2}
                                \opn{Im} (s a^*)
                                r (r^2 - 1)
                                \frac{1}{m_A m_B} 
                                \quad
                            &
                                ( m_A < m_B = m_S ) \,,
                                \\
                            &
                                \frac{1}{2 (4 \pi)^2}
                                \opn{Im} (s a^*)
                                r (r^2 - 1)
                                \frac{m_B}{m_A^3} 
                                \qquad 
                            &
                                ( m_B < m_A = m_S )\,,
                                \\
                            &
                                \frac{1}{12 (4 \pi)^2}
                                \opn{Im} (s a^*)
                                r (r^2 - 1)
                                \frac{1}{m_A m_B} 
                                \qquad 
                            &
                                ( m_S < m_A = m_B ) \,.
                        \end{aligned}
                    \right.
                \label{eq:Weinberg r1r}
            \end{aligned}
        \end{align}
In the case of $(r,r,1)$, the induced Weinberg operator is symmetric under $m_A$ and $m_{B}$, so that we omit the case of $m_B < m_A = m_S$.

The electroweak-Weinberg operator is a dimension-six operator so that the mass dimension of $C_W$ is $-2$; $C_W$ scales as  $1/(m_A m_B)$ in Eqs.~\eqref{eq:Weinberg rr1} and \eqref{eq:Weinberg r1r}
when $m_A<m_B$ or $m_B<m_A$  except for the case of $(r,1,r)$. This behavior is expected in the effective theory description, as discussed in Sec.~\ref{sec:sun_weinberg}. When the lighter fermion is SU(2)$_L$ non-singlet, it has the SU(2)$_L$ EDM after the heavier fermion is integrated out. The electroweak-Weinberg operator, generated by integrating out the lighter fermion, is proportional to the SU(2)$_L$ EDM as in Eq.~\eqref{eq: d_A_to_C_W}. When the lighter fermion $\psi_B$ is SU(2)$_L$ singlet in the case of $(r,1,r)$, $C_W$ does not get such a contribution. Then, it is suppressed by $m_B/m_A^3$ when $m_B <m_A=m_S$. When $m_S<m_A=m_B$, $C_W$ is proportional to $1/(m_Am_B)$, not enhanced by $1/m_S$. In this case, after integrating out the fermions, the scalar is left in the effective theory.  The electroweak-Weinberg operator does not get any contribution by integrating out the scalar even if the scalar is SU(2)$_L$ non-singlet. 

\begin{figure}
            \begin{tabular}{c}
            \vspace{0.3cm}
                \begin{subfigure}{1\textwidth}
                    \centering
                    \includegraphics[width=\textwidth]{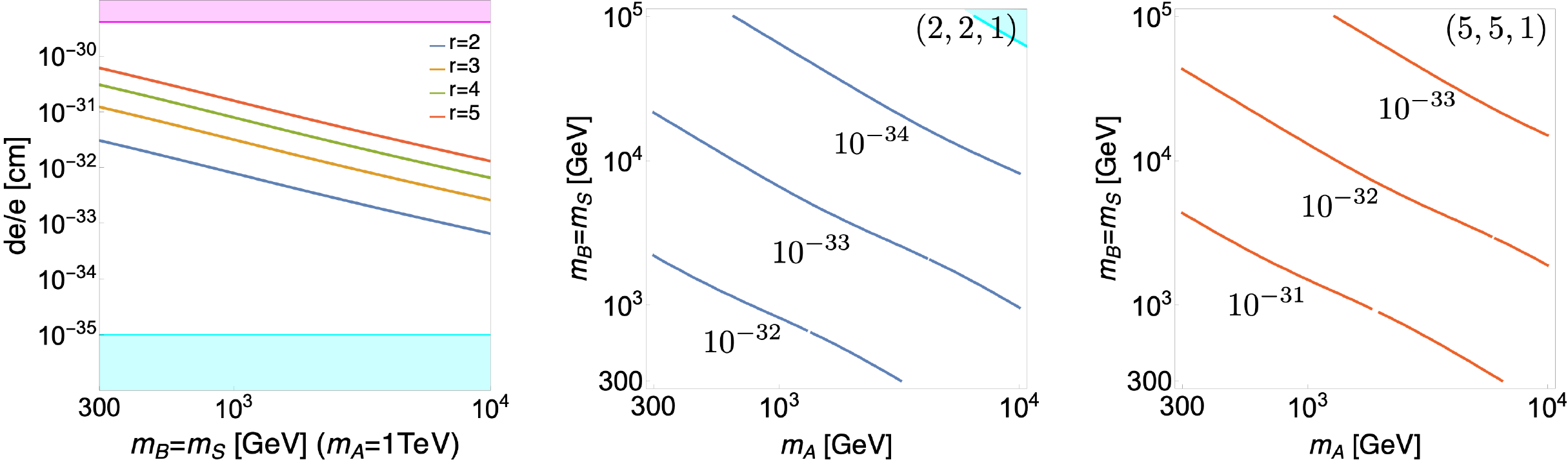}
                    \caption{                        Under the assumption of $m_B=m_S$.  
                    }
                    \label{fig:Weinberg rr1A}
                \end{subfigure}
                \\
                \begin{subfigure}{1\textwidth}
                    \centering
                    \includegraphics[width=\textwidth]{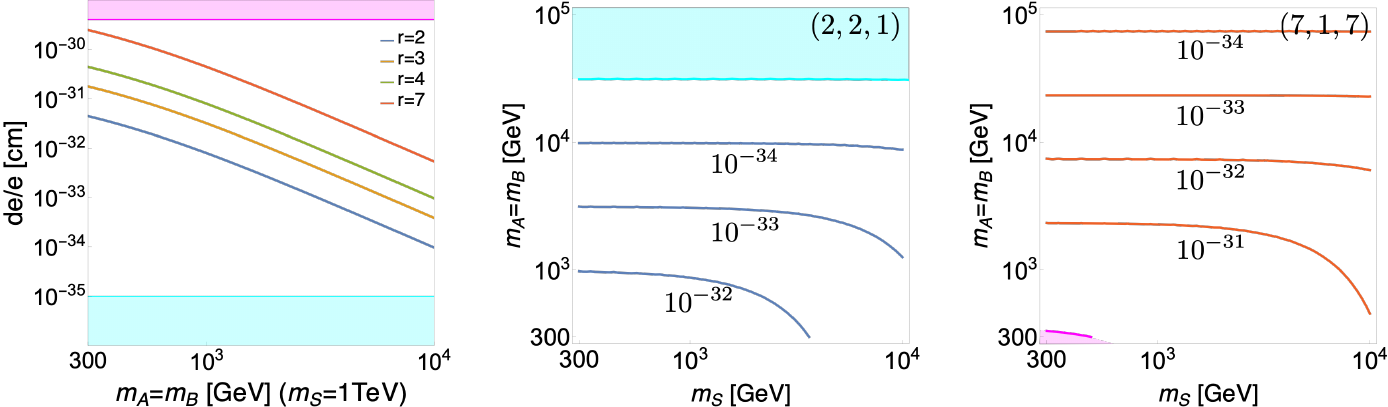}
                    \caption{
Under the assumption of $m_A=m_B$. 
                    }
                    \label{fig:Weinberg rr1S}
                \end{subfigure}
            \end{tabular}
            \caption{
                Electron EDM induced by the Yukawa coupling of the $(r,r,1)$ SU(2)$_L$ multiplets. Here,  $\opn{Im} (s a^*) = 0.25$ is taken.  In the upper (lower) panels, $m_B=m_S$ ($m_A=m_B$) is assumed. In the left panels, the four color lines illustrate each SU(2)$_L$ representation, where $r=2,3,4,5$, respectively. The SU(2)$_L$ representations in the middle and right panels are shown inside the figures.        
                The magenta bands denote the experimental bound on the electron EDM $|d_e^{\rm exp.}| < 4.1 \times 10^{-30} e \, {\rm cm}$ \cite{Roussy:2022cmp}, and the cyan-shaded regions can not be probed due to the CKM contribution through the $e$--$N$ four-fermion interaction~\cite{Ema:2022yra}.
            }
            \label{fig:Weinberg rr1}
        \end{figure}

\begin{figure}
            \centering
            \begin{tabular}{c}
            \vspace{0.3cm}
                \begin{subfigure}{1\textwidth}
                    \includegraphics[width = \textwidth]{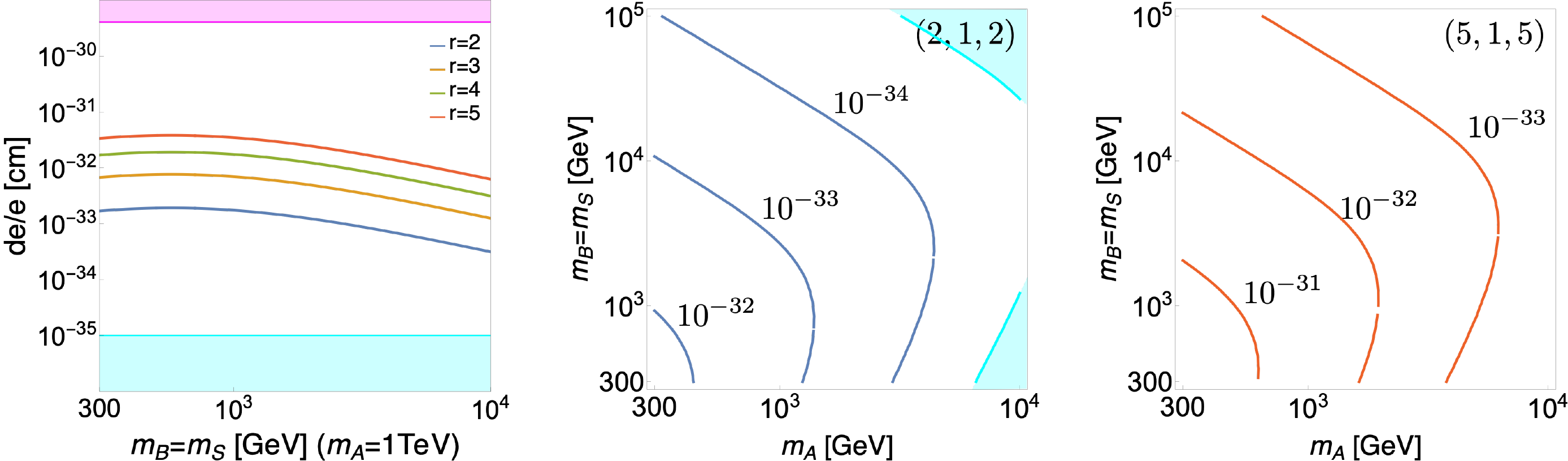}
                    \caption{
Under the assumption of $m_B=m_S$.
                    }
                    \label{fig:Weinberg r1rA}
                \end{subfigure}
                \\
                \vspace{0.3cm}
                \begin{subfigure}{1\textwidth}
                    \includegraphics[width = \textwidth]{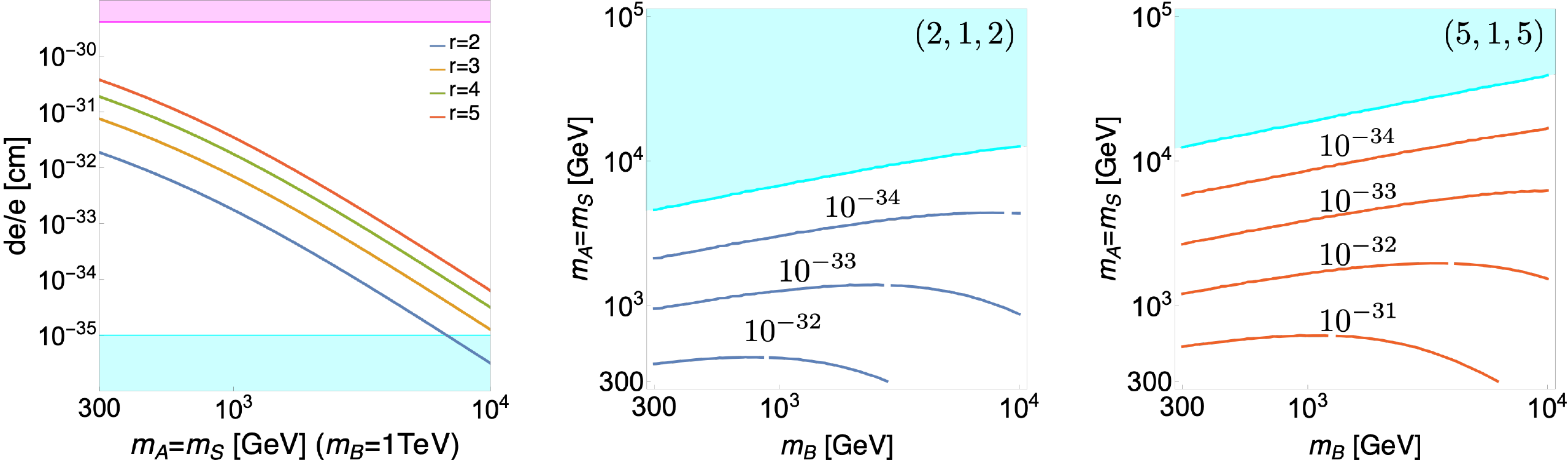}
                    \caption{
Under the assumption of $m_A=m_S$.
                    }
                    \label{fig:Weinberg r1rB}
                \end{subfigure}
                \\
                \begin{subfigure}{1\textwidth}
                    \includegraphics[width = \textwidth]{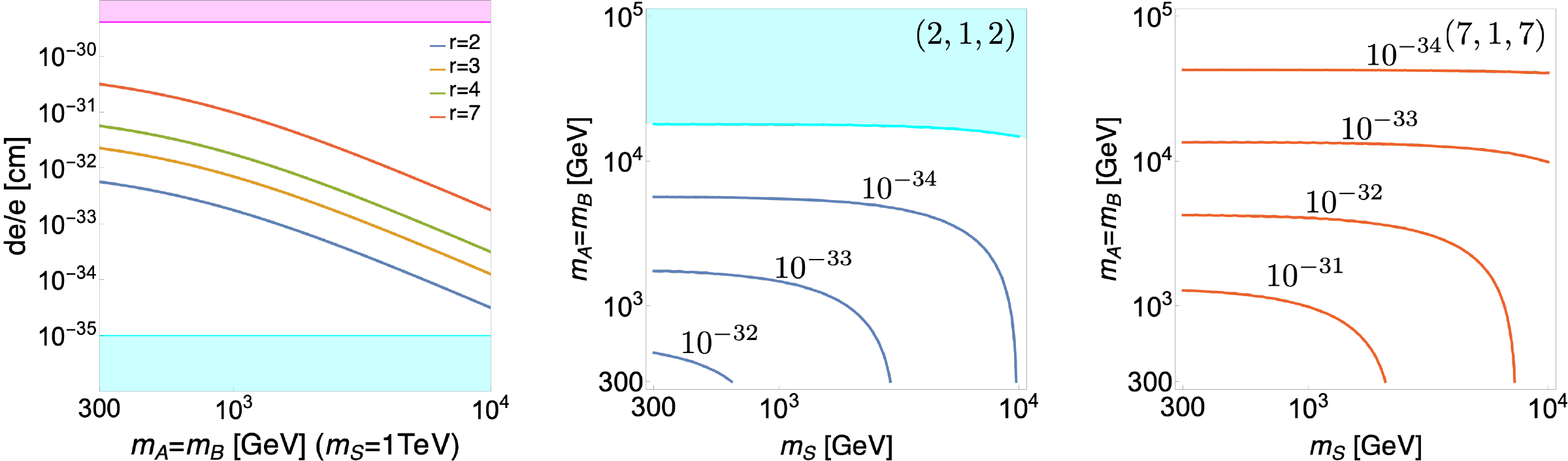}
                    \caption{
Under the assumption of $m_A=m_B$.
                    }
                    \label{fig:Weinberg r1rS}
                \end{subfigure}
            \end{tabular}
            \caption{
                Electron EDM induced by the Yukawa coupling of the $(r,1,r)$ SU(2)$_L$ multiplets. In the upper, middle, and lower panels, 
                $m_B=m_S$ $m_A=m_S$, and $m_A=m_B$ are assumed, respectively. The others are the same as in Fig.~\ref{fig:Weinberg rr1}.
                           }
            \label{fig:Weinberg r1r}
        \end{figure}

Equations~\eqref{eq:Weinberg rr1} and \eqref{eq:Weinberg r1r} tell us that the electron EDM grows proportionally to $r (r^2-1)$. $C_W$ is largely enhanced when the dimension of SU(2)$_L$ multiplets $r$ is larger.  In the minimal dark matter models, large representations in SU(2)$_L$ are introduced for the dark matter stability; 
$r=5$ for the fermionic dark matter and $r=7$ for the scalar dark matter \cite{Cirelli:2005uq}.
The numerical analyses are shown in Figs.~\ref{fig:Weinberg rr1} and \ref{fig:Weinberg r1r}, and the above enhancement proportional to $r (r^2-1)$ is shown there.
Figure~\ref{fig:Weinberg rr1} is drawn under the case of the representation $(r,r,1)$.  In the left panel of Fig.~\ref{fig:Weinberg rr1A},   
we assume $m_A=1$\,TeV and $m_B=m_S$. The four color lines illustrate each SU(2)$_L$ representation, where $r=2,3,4,5$, respectively.
We fix the Yukawa coupling at $\opn{Im} (s a^*) = 0.25$ and use the value in the following figures. Other physical constants required are the electron mass and the SU(2)$_L$ coupling constant $\alpha_2 = 0.034$ \cite{PDG:2024}.
Here, we take $m_B=m_S>300$~GeV so that the effective theory description, including the electroweak-Weinberg operator, works after integrating out the heavy particles. The magenta region is excluded by the current experimental bound on the electron EDM ($|d^{\rm exp}_e| < 4.1 \times 10^{-30}e$\,cm) \cite{Roussy:2022cmp}. 
In the cyan-shaded region, the paramagnetic atom or molecule EDMs get the dominant contribution from the CKM phase by the semi-leptonic four-Fermi operators, so it is difficult for the future measurements of the electron EDM to discover the BSM contribution to the electron EDM  smaller $d^{\rm eq}_e=1.0\times 10^{-35}e$\,cm \cite{Ema:2022yra}.

In the middle of Fig.~\ref{fig:Weinberg rr1A}, we show the electron EDM in the case of $(2,2,1)$ as a function of $m_A$ and $m_B=m_S$. Even when $m_A = 1$ TeV and $m_B = m_S = 10$ TeV with $r=2$, the electron EDM reaches $|d_e^{(2,2,1)}| \simeq 2.2 \times 10^{-34} e \, {\rm cm}$, larger than $d^{\rm eq}_e$. 
The right panel is for the case of $r=5$. The $r=5$ fermion is introduced in a minimal dark matter model \cite{Cirelli:2005uq}. The electron EDM is 20 times larger than $r=2$. The thermal relic abundance of the dark matter favors the mass of the $r=5$ fermion to be below 10\,TeV for $\Omega_{\rm DM}h^2 \leq 0.11$ \cite{Cirelli:2007xd,Cirelli:2009uv} after including the Sommerfeld effect \cite{Hisano:2006nn}. Even such a heavy mass might be accessible in future electron EDM measurements.  

In the lower panels in Fig.~\ref{fig:Weinberg rr1}, we take the same parameters as the upper panels except for assuming $m_A = m_B$. The electron EDM is insensitive to $m_S$ as far as $m_S\lesssim m_A=m_B$ as in Eq.~\eqref{eq:Weinberg r1r}. On the other hand, when $m_S\gg m_A=m_B$, the EDM is suppressed by $1/m_S^2$ since the SU(2)$_L$ EDMs for $\psi_A$ and $\psi_B$ are also suppressed, as discussed in Sec.~\ref{sec:sun_weinberg}. It is found that these figures imply the models are highly expected to be explored by the improved experimental results in a few decades.
        
Next, Fig.~\ref{fig:Weinberg r1r} shows the scenario of $(r,1,r)$. 
In Figs.~\ref{fig:Weinberg r1rA}, \ref{fig:Weinberg r1rB}, and 
\ref{fig:Weinberg r1rS}, we assume $m_B=m_S$, $m_A=m_S$, and $m_A=m_B$, respectively. The left panels of those figures have different behaviors with varying heaver particle masses. They are scaled as $1/(m_A m_B)$, $m_B/m_A^3$, and $1/(m_A m_B)$, respectively. This is expected from the effective theory description as discussed below Eq.~\eqref{eq:Weinberg r1r}.

We show contour plots with the cases of  $(2,1,2)$ and $(5,1,5)$
Figs.~\ref{fig:Weinberg r1rA} and  \ref{fig:Weinberg r1rB}, while those of  $(2,1,2)$ and $(7,1,7)$ are shown in Fig.~\ref{fig:Weinberg r1rS}. The scalar of $r=7$ is introduced in the minimal dark matter models, and the mass is favored to be 25\,TeV from the thermal relic abundance \cite{Cirelli:2007xd}. 
The electron EDM is 73.5 times larger than $r=2$. It depends on the masses of fermions coupled with the scalar, not the scalar mass itself.

\begin{figure}
            \centering
            \begin{tabular}{c}
            \vspace{0.3cm}
                \begin{subfigure}{1\textwidth}
                    \centering
                    \includegraphics[width = 0.8\textwidth]{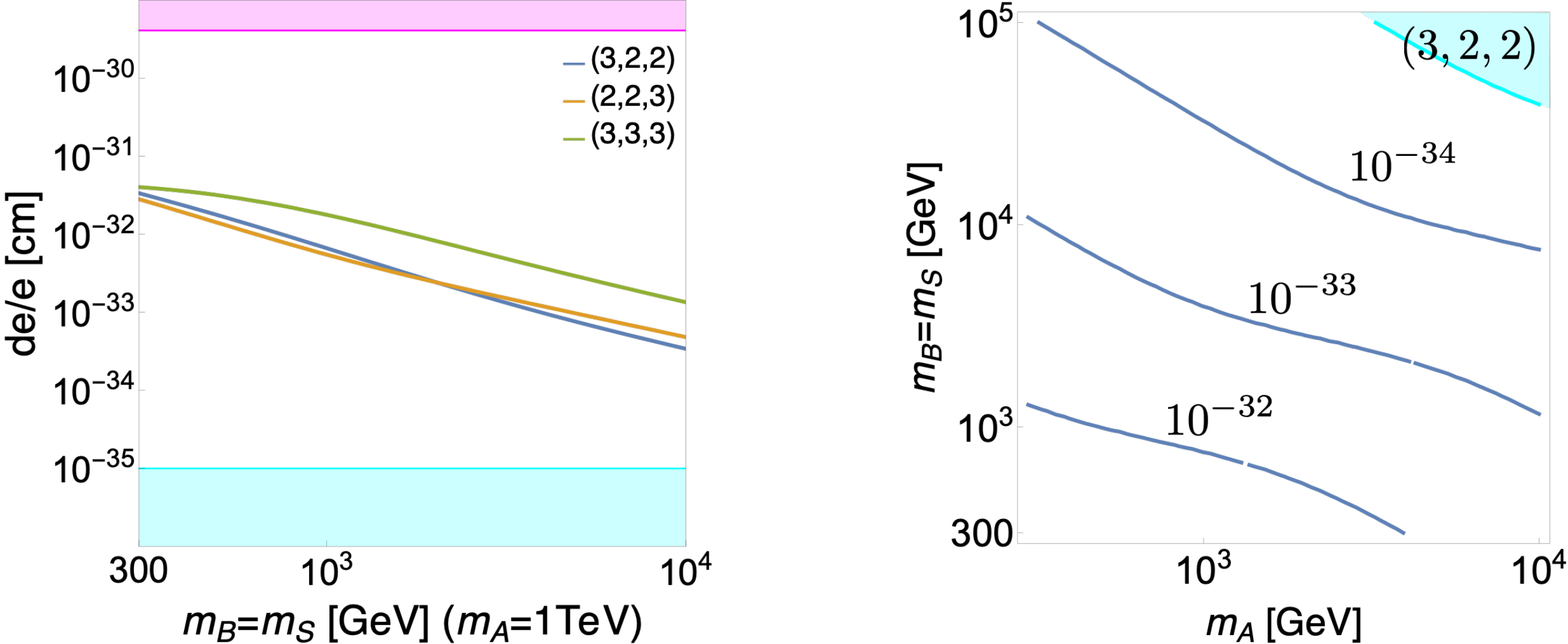}
                    \caption{
                         Under the assumption of  $m_B = m_S$.
                    }
                    \label{fig:Weinberg S 3A}
                \end{subfigure}
                \\
                \vspace{0.3cm}
                \begin{subfigure}{1\textwidth}
                    \centering
                    \includegraphics[width = 0.8\textwidth]{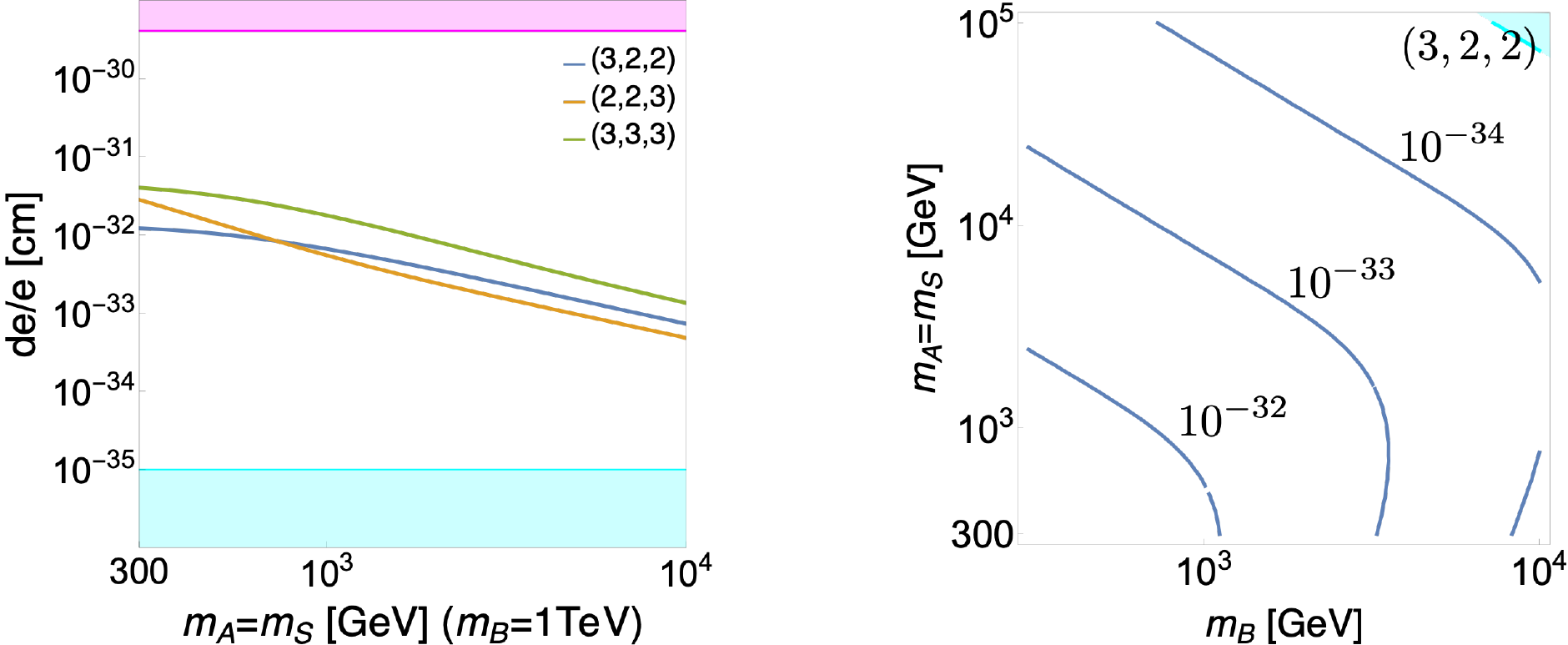}
                    \caption{
                        Under the assumption of $m_A = m_S$.
                    }
                    \label{fig:Weinberg S 3B}
                \end{subfigure}
                \\
                \begin{subfigure}{1\textwidth}
                    \centering
                    \includegraphics[width = 0.8\textwidth]{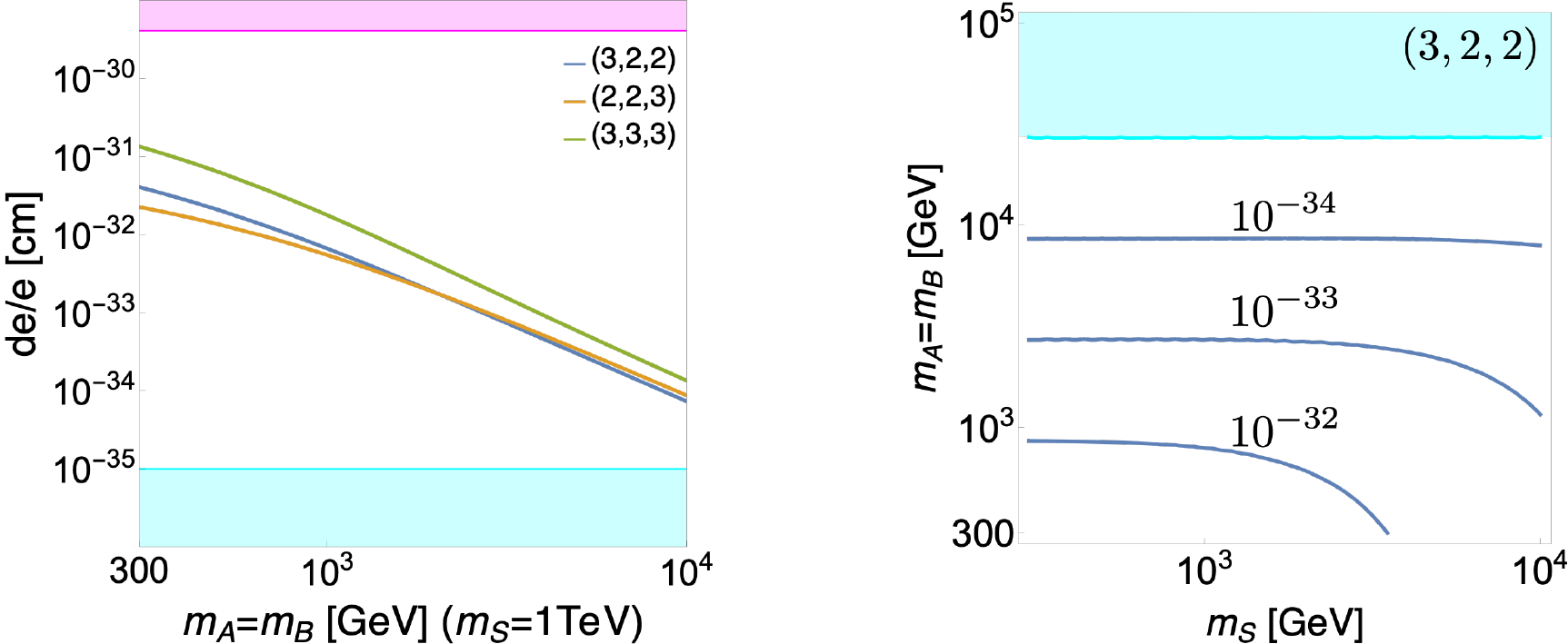}
                    \caption{
                         Under the assumption of  $m_A = m_B$.
                    }
                    \label{fig:Weinberg S 3S}
                \end{subfigure}
            \end{tabular}
            \caption{
                Electron EDMs induced by the electroweak-Weinberg operator, under the assumption that the SU(2)$_L$ representations are  $(A,B,S) = (3,2,2)$, $(2,2,3)$ and $(3,3,3)$.
                The others are the same as in Fig.~\ref{fig:Weinberg rr1}.
            }
            \label{fig:Weinberg S 3}
        \end{figure}

Finally, we show results for the cases of $(3,2,2), \, (2,2,3)$ and $(3,3,3)$ in Fig.~\ref{fig:Weinberg S 3}. The group factors in the formula of $C_S$ are derived from Table~\ref{tab:group_factors_SU2}. Since the fermions are SU(2)$_L$ non-singlet, the electron EDM (and also $C_W$) is scaled as $1/(m_A m_B)$ as far as $m_S$ is not much heavier than $m_A$ and $m_B$.


\subsection{$S = H$}
Next, we discuss the importance of the electroweak-Weinberg operator contribution in the case that the scalar $S$ is identified with the SM Higgs. It is assumed that the Higgs has the CP-violating Yukawa couplings with $\psi_A$ and $\psi_B$. In this case, the two-loop Barr-Zee diagram is the dominant contribution to the electron EDM. 
The radiative correction to the contribution might be comparable to that from the electroweak-Weinberg operator, both of which are three-loop level contributions.
We will clarify this point first before showing the numerical results.

Let us assume $m_S(=m_H)\ll m_A<m_B$. Integrating out $\psi_B$ at tree level generates the dimension-five operator of the Higgs boson and $\psi_A$, and then integrating out $\psi_A$ at one-loop level induces the effective operator $|H|^2 F^{\mu\nu}\widetilde{F}_{\mu\nu}$.
The coefficient is suppressed by $1/(m_Am_B)$. Another loop diagram with the effective operator induces the electron EDM. As a result, the electron EDM is proportional to $m_e/(m_Am_B)$ with a two-loop factor. This 
implies that the electroweak-Weinberg operator contribution has mass parameter dependence similar to the Barr-Zee contribution. In Ref.~\cite{Kuramoto:2019yvj}, the anomalous dimensions for the dimension-five operator of the Higgs boson are evaluated at one-loop level. The corrections to the electron EDM from the anomalous dimensions might be comparable to the contribution from the electroweak-Weinberg operator.

For concreteness, we assume $(A,B,S)=(3,2,2^H)$, where $2^H$ means the SM Higgs boson.
The Wilson coefficients $C_W$ for the electroweak-Weinberg operator are  approximately given as  
\begin{align}
     &\begin{aligned}
        C_W
        &\simeq \left\{
\begin{aligned}
& \frac{1}{(4\pi)^4}\Im(sa^*)
            \times
            \frac{m_A}{8m_B^3}\left(11+8\ln\frac{m^2_A}{m^2_B}\right)
 \quad
 &( m_H \ll m_A < m_B ) \,,
                                \\
&
\frac{1}{(4\pi)^4} \Im(sa^*)
            \times
                \frac{-5}{8m_Am_B}
\quad
 &( m_H \ll m_B < m_A ) \,.
                        \end{aligned}
                    \right.
\label{eq:Weinberg 3^l22}
\end{aligned}
\end{align}
Here, we take the triplet fermion $\psi_A$ with the Majorana mass $m_A$ and the doublet fermion $\psi_B$ with Dirac mass $m_B$.
When $m_B<m_A$, $C_W$ is proportional to $1/(m_A m_B)$. On the other hand, when $m_A<m_B$, it is suppressed by $m_A/m_B^3$. It comes from accidental cancellation between the leading contributions of $g_1$ and $g_2$ in $C_W$ in a limit of $m_H\rightarrow 0.$\footnote{
It can be directly checked by using Eq.~\eqref{eq:appriximationofg1andg2}.}
Thus, the contribution to the electron EDM is more suppressed in the latter case. 

Now we take a ratio between the contributions from the electroweak-Weinberg operator ($d_e^{(W)}$) and from the correction to the Barr-Zee diagrams (${\delta d_{e}^{(BZ)}}$). It is approximately given as 
    \begin{align}
            \frac{d_e^{(W)}}{\delta d_{e}^{(BZ)}}
        &\simeq \left\{
\begin{aligned}
&              \frac{1}{192}
            \frac{\alpha^2_2}{\alpha_e}
            \left(
                \bar{\gamma}^{(3,0)}_{ss}
                \ln \frac{m_B}{m_A}
            \right)^{-1}
            \frac{m^2_A}{m_B^2}
            \frac{
                11
                +
                8 \ln \frac{
                    m^2_A}{m^2_B
                }
            }{
                2
                +
                \ln \frac{m^2_A}{m_H^2}
            }
 \quad
& ( m_H \ll m_A < m_B ) \,,
                                \\
& \frac{5}{48}
            \frac{\alpha^2_2}{\alpha_e}
            \left[
                \left(
                    3\bar{\gamma}^{(2,1/2)}_{ss}
                    +
                    \bar{\gamma}^{(2,1/2)}_{tt}
                \right)
                \ln \frac{m_A}{m_B}
            \right]^{-1}        
            \frac{1}{
                2+\ln \frac{m^2_B}{m_H^2}
            } 
\quad
 &( m_H \ll m_B < m_A ) \,,
\end{aligned}
                    \right.
\label{eq:Weinberg 3^l22}
\end{align}
where $\bar{\gamma}^{(r,Y_r)}_{ij}/(4\pi)\;(i,j=s,t)$ are the components of anomalous dimension matrix for the dimension-five operators of $\psi \psi H^2$ for the Barr-Zee diagrams. 
Depending  on the gauge charges of the lighter fermion, they  
are given as \cite{Bishara:2018vix,Kuramoto:2019yvj}
\begin{align}
    \bar{\gamma}^{(r,Y_r)}_{ss}&=-\left[6\alpha_2\left(C_2(r)+\frac34 \right)
    +6\alpha_Y\left(Y_r^2+\frac14\right)
    -3\lambda'-6\alpha_t\right]\,,\\
    \bar{\gamma}^{(r,Y_r)}_{tt}&=-\left[6\alpha_2\left(C_2(r)-\frac14 \right)+6\alpha_Y\left(Y_r^2+\frac14 \right)-\lambda'-6\alpha_t\right]\,,
\end{align}
where $\alpha_2, \alpha_Y, $ and $\alpha_t$ are for SU(2)$_L$, U(1)$_Y$ and top-Yukawa coupling constants, respectively, and $\lambda'(=\lambda/4\pi)$ is for the SM Higgs quartic coupling constant ($\lambda$). $C_2(r)$ and $Y_{r}$ are the Casimir operator and the hypercharge for the lighter fermion. ($C_2=2$ and $Y_r=0$ for $\psi_A$ and $C_2=3/4$ and $Y_r=1/2$ for $\psi_B$.)

    \begin{figure}
            \centering
            \begin{tabular}{cc}
                \begin{subfigure}{0.48 \textwidth}
                    \includegraphics[width = \textwidth]{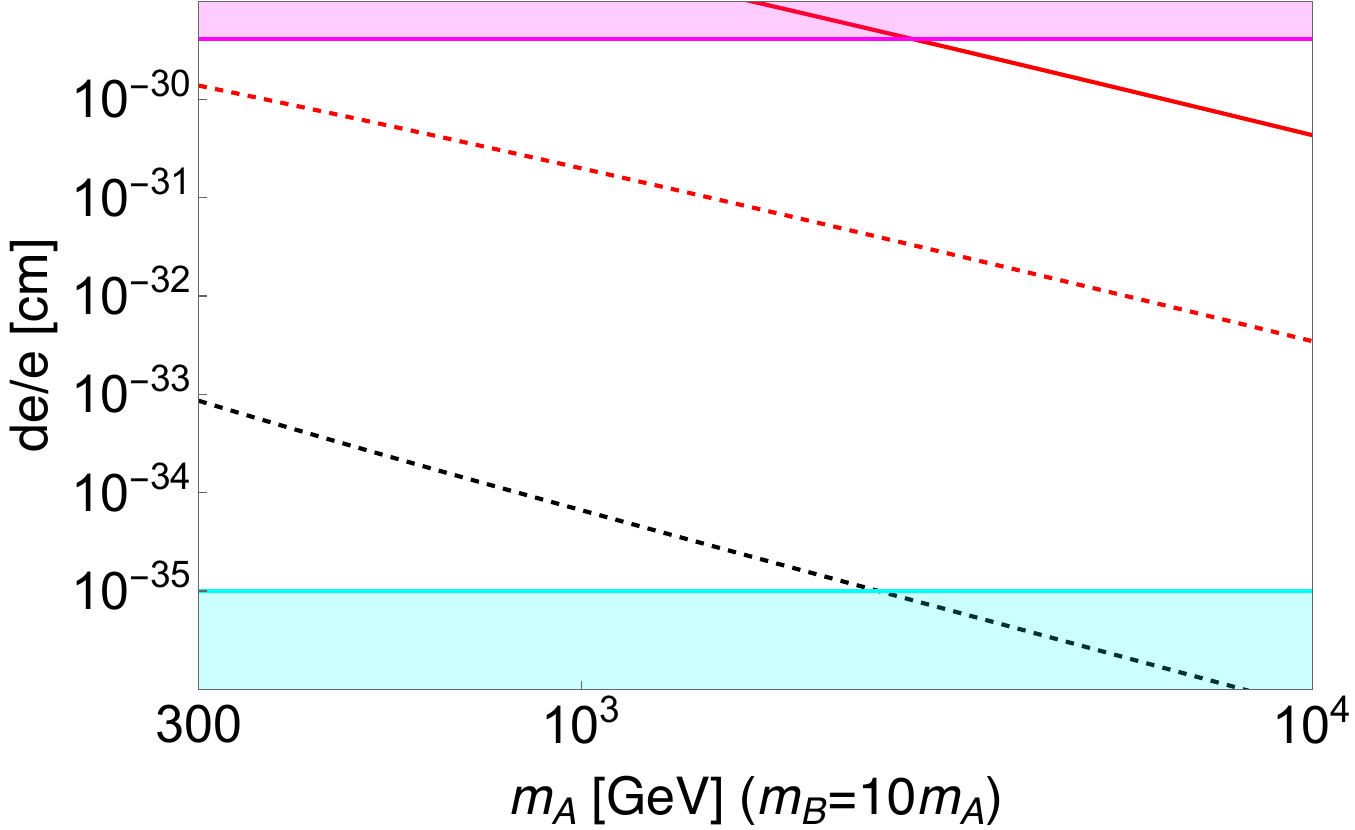}
                    \caption{
                    $m_B=10m_A$    
                    }
                    \label{fig:Weinberg H 3A}
                \end{subfigure}
                &
                \begin{subfigure}{0.48 \textwidth}
                    \includegraphics[width = \textwidth]{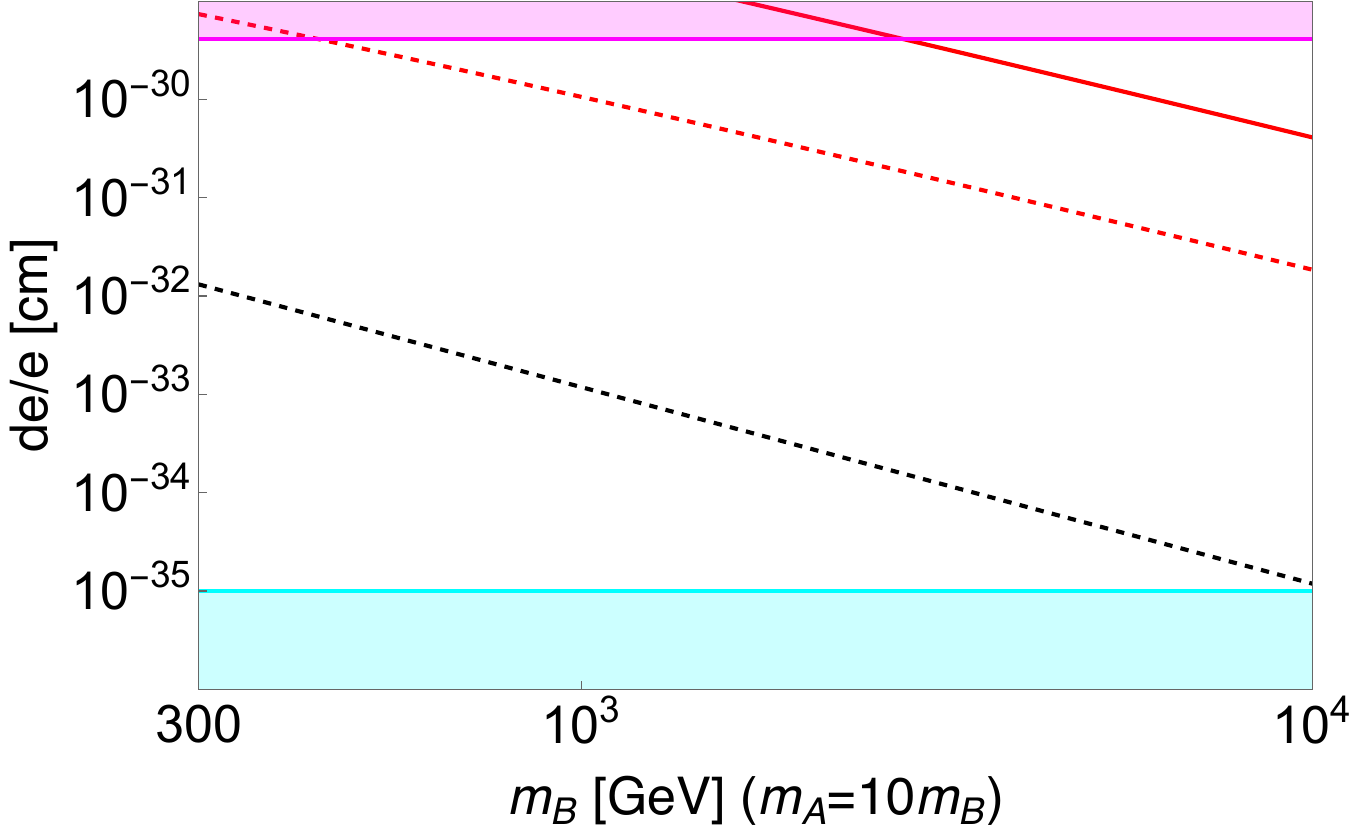}
                    \caption{
                    $m_A=10m_B$ 
                    }
                    \label{fig:Weinberg H 2B}
                \end{subfigure}
            \end{tabular}
            \caption{Contributions to the electron EDM from the electroweak-Weinberg operator (dotted-black lines) and the radiative correction in the Barr-Zee diagrams (the dotted-red lines). 
            The red-solid lines are for the sum of LO and NLO contributions in the Barr-Zee diagrams. We assume $(A,B,S) = (3,2,2^H)$ ($2^H$ stands for the SM Higgs). In the left (right) panel, the triplet Majorana fermion (doublet Dirac fermion) is 10 times lighter than the doublet (triplet) one.  $\opn{Im} (s a^*) = 0.25$ is taken.
                }
            \label{fig:Weinberg H 3}
    \end{figure}  

As expected, the ratio in Eq.~\eqref{eq:Weinberg 3^l22} is not suppressed by the power of the coupling constants. It is suppressed by $m^2_A/m^2_B$ in 
the case of $m_A<m_B$, while it is not in $m_A>m_B$. However, 
we find that the ratio is suppressed numerically in both cases. 
In Fig.~\ref{fig:Weinberg H 3}, we show the contributions to the electron EDM from the correction to the Barr-Zee diagrams (${\delta d_{e}^{(BZ)}}$, the dotted red lines) and from the electroweak-Weinberg operator ($d_e^{(W)}$, the dotted black lines), and the Barr-Zee diagram contribution including the correction (the red solid line). The Figs.~\ref{fig:Weinberg H 3A} and ~\ref{fig:Weinberg H 2B} correspond to the cases of $m_A<m_B$ and  $m_B<m_A$, respectively.  Here, the heavier fermion mass is taken to be 10 times larger than the lighter one ($m_B=10 m_A$ or $m_A=10 m_B$).
We take $m_t=172.57\;\rm GeV$, and $\alpha_Y=\alpha_2 \tan^2 \theta_w=0.01$~\cite{PDG:2024} as input parameters. We found that {$d_e^{(W)}/{\delta d_{e}^{(BZ)}}$ is $3.3\times 10^{-4}\;(1.1\times 10^{-3})$} for $m_A<m_B$ ($m_B < m_A$)
when the lighter fermion mass is 1\,TeV. 

\section{Conclusions and discussion}
\label{sec:conclusion}

The progress of the electron EDM measurements has been remarkable. The CP-violating interactions induced by BSM around the TeV scale have been constrained even if the electron EDM is generated at two-loop level. In the coming decades,
the experimental improvements are expected to probe the contributions even at three-loop level. 

In this paper,
we study the electron EDM generated by the electroweak-Weinberg operator at three-loop level.
If the CP-violating Yukawa couplings are introduced with SU(2)$_L$ BSM multiplets, the electroweak-Weinberg operator is generated at two-loop level. Below the electroweak scale,  
the electron EDM is radiatively induced at three-loop level since it is generated by another one-loop diagram with the electroweak-Weinberg operator interaction. 
We introduce the SU(2)$_L$ multiplets with TeV scale masses, which could be motivated by the dark matter multiplets,
and investigate the predicted size of the electron EDM. 
It is found that the future electron EDM measurements would cover the prediction and might discover them since it may be larger than the SM contributions to the paramagnetic atom or molecule EDMs, $d_e^{\rm eq}=1.0\times 10^{-35}e$\,cm. 
We notice that if large SU(2)$_L$ multiplets, such as five- or seven-dimensional multiplets, are introduced, the electron EDM is enhanced by the cubic power of the dimension. Such large-dimensional multiplets are motivated in the minimal dark matter models due to the stability of dark matter. 

We also discuss the relation between the Barr-Zee diagram and the electroweak-Weinberg operator contributions to the electron EDM. 
If the SM Higgs has 
the CP-violating Yukawa coupling with the SU(2)$_L$ BSM multiplets, the Barr-Zee diagrams at two-loop level contribute to the electron EDM. Thus, the radiative correction to the Barr-Zee diagrams might be comparable to the contribution from the electroweak-Weinberg operator. We compute the radiative correction to the Barr-Zee diagrams using the anomalous dimensions for the dimension-five operators of the SM Higgs and the fermion generated by integrating out the heavier fermion, and compare it with that from the electroweak-Weinberg operator contribution.
We find the electroweak-Weinberg operator contribution is numerically smaller than the radiative correction to the Barr-Zee diagrams, and it can be safely negligible. 

Our evaluation of the electron EDM can be improved further. We ignored contributions from the evanescent operators for the electroweak-Weinberg operators, though we evaluated the matching condition of the operators to the electron EDM at one-loop level. In addition, we did not also evaluate the contribution from the CP-violating SU(2)$_L$ dipole moment operator of lepton in the SMEFT. They may contribute to the electron EDM,  comparably to the electroweak-Weinberg operators.

We did not evaluate the electroweak-Weinberg operator contribution to the light quark EDMs, though the calculation is straightforward. They contribute to the hadronic EDMs, such as the neutron and Mercury ones.  However, the electron EDM constrains the electroweak-Weinberg operator much more severely, and the hadronic EDM measurements are not competitive with the electron EDM even in their future prospects.\footnote{
The hadronic EDM measurements are still important to constrain the QCD $\theta$ parameter.
}

\acknowledgments

We would like to thank Ryosuke Sato for the useful discussion.
This work is supported by the JSPS Grant-in-Aid for Scientific Research Grant No.\,23K20232 (J.H.),  No.\,24K07016 (J.H.), No.\,24K22872 (T.K.), and No.\, 24KJ1256 (N.O.).
The work of J.H.\ is also supported by 
World Premier International Research Center Initiative (WPI Initiative), MEXT, Japan.
This work is supported by 
JSPS Core-to-Core Program Grant No.\,JPJSCCA20200002. 
The work of T.B.\ was financially supported by JST SPRING, Grant Number JPMJSP2125.
The author T.B.\ would like to take this opportunity to thank the ``THERS Make New Standards Program for the Next Generation Researchers".

\bibliographystyle{utphys28mod}
\bibliography{ref}

\end{document}